\documentclass[useAMS]{mn2e}

\usepackage{ulem} 
\usepackage{graphicx}
\usepackage{ulem}
\usepackage{rotating}

\newcommand{\hst}{{\sl HST}}

\def\lapprox{\hbox{\lower .8ex\hbox{$\,\buildrel < \over\sim\,$}}}
\def\gapprox{\hbox{\lower .8ex\hbox{$\,\buildrel > \over\sim\,$}}}

\title[Improved {\it HST\/} Proper Motion for Tycho-G]{Improved {\it Hubble 
Space Telescope} Proper Motions for Tycho-G and Other Stars in the
Remnant of Tycho's Supernova 1572
}
\author[L.\ R.\ Bedin et al.]{ 
L.\ R.\ Bedin$^{1}$, 
P.\ Ruiz-Lapuente$^{2,3,4}$, 
J.\ I.\ Gonz\'alez Hern\'andez$^{5,6}$,  
R.\ Canal$^{7,3}$, 
\newauthor
A.\ V. Filippenko$^{8}$,
J.\ Mendez$^{9,7}$
\\  
$^{1}$INAF-Osservatorio Astronomico di Padova, Vicolo
             dell'Osservatorio 5, I-35122 Padova, Italy\\
$^{2}$Instituto de F\'{\i}sica Fundamental, Consejo Superior de
       Investigaciones Cient\'{\i}ficas, c/Serrano 121, E--28006 Madrid, 
       Spain\\ 
$^{3}$Institut de Ci\`encies del Cosmos (UB--IEEC), c/ Mart\'{\i} i 
       Franqu\'es 1, E--08028 Barcelona, Spain, E-mail: pilar@am.ub.es\\ 
$^{4}$ Max-Planck Institut f\"ur Astrophysik
       Karl Schwarzschildstrasse 1
       D 85748 Garching Germany\\
$^{5}$Instituto de Astrof\'{\i}sica de Canarias, E--38205 La Laguna, 
       Tenerife, Spain\\ 
$^{6}$Departamento de Astrof\'{\i}sica, Universidad de La Laguna, E--38206 
       La Laguna, Tenerife, Spain\\ 
$^{7}$Departament d'Astronomia i Meteorologia, Universitat de Barcelona, 
       c/ Mart\'{\i} i Franqu\'es 1, 08028 Barcelona, Spain\\
$^{8}$Department of Astronomy, University of California, Berkeley, 
       CA 94720--3411, USA\\
$^{9}$Isaac Newton Group of Telescopes, PO Box 321, E--38700 Santa Cruz 
       de la Palma, Spain\\ 
}
\begin{document}

\date{Accepted 2013 December 18. Received 2013 December 18; in original form 
2013 May 6}

\pagerange{\pageref{firstpage}--\pageref{lastpage}} \pubyear{2013}

\maketitle

\label{firstpage}
 
\begin{abstract}
With archival and new \textit{Hubble Space Telescope\/} observations
we have refined the space-velocity measurements of the stars in the
central region of the remnant of Tycho's supernova (SN) 1572, one of
the historical Galactic Type Ia supernova remnants (SNRs).
We derived a proper motion for Tycho-G of
$(\mu_{\alpha\cos{\delta}};\mu_\delta)_{\rm J2000.0}= (-2.63;-3.98)
\pm (0.06;0.04)$ [formal errors] $\pm$ $(0.18;0.10)$ [expected errors] mas yr$^{-1}$.
We also reconstruct the binary orbit that Tycho-G should have followed
if it were the surviving companion of SN~1572.
We redetermine the Ni abundance of this star and compare it with new
abundance data from stars of the Galactic disk, finding that [Ni/Fe]
is about 1.7\,$\sigma$ above the Galactic trend.
From the high velocity ($v_{b} = -50\pm14$ km s$^{-1}$) of
Tycho-G perpendicular to the Galactic plane, its metallicity, and its
Ni excess, we find the probability of its being a chance interloper to
be $P \lapprox 0.00037$ at most.
The  projected rotational velocity of the star should be below current 
observational limits.
The projected position of Tycho-G is, within the uncertainties,
consistent with the centroid of the X-ray emission of Tycho's SNR;
moreover, its brightness is generally consistent with the
post-explosion evolution of the luminosity of a SN companion.
Among the other 23 stars having $V < 22$ mag and located within $42''$
from the X-ray centroid, only 4 are at distances compatible with that
of the SNR, and none of them shows any peculiarity.  
Therefore, if even Tycho-G is {\it not} the surviving companion of SN~1572, 
the absence of other viable candidates does favor the merging
of two white dwarfs as the producer of the SN. 
\end{abstract}


\section{Introduction}
\label{introduction}

Type Ia supernovae (SNe~Ia) have long been recognised as being close
binary systems where one of the stars, a carbon-oxygen white dwarf
(C+O WD), undergoes a thermonuclear runaway after reaching explosive
conditions at its centre (e.g., Branch et al.\ 1995, and references
therein). The physics of the explosion is determined by both
components of the system. While the ejecta left by the explosion have
been studied in great detail, the direct search for the companion is a
relatively new approach (Ruiz-Lapuente 1997) that has already started
to give results (Ruiz-Lapuente et al.\ 2004, hereafter RL04;
Gonz\'alez Hern\'andez et al.\ 2009, hereafter GH09; Schaefer \&
Pagnotta 2012; Edwards et al.\ 2012; Gonz\'alez Hern\'andez et
al.\ 2012; Kerzendorf et al.\ 2012).

SNe~Ia are the best cosmological distance indicators, and they were
used to discover the accelerating expansion of the Universe (Riess et
al.\ 1998; Perlmutter et al.\ 1999).  While they are well
characterised for empirical use as cosmological probes, there are
still significant gaps in our theoretical understanding of them.
%
%
In principle, SNe~Ia can be produced through two different channels:
the single-degenerate (SD) channel and the double-degenerate (DD)
channel (Whelan \& Iben\ 1973; Webbink\ 1984; Iben \& Tutukov\ 1984;
Livio \& Truran\ 1992; Branch et al.\ 1995; Ruiz-Lapuente\ 1997;
Livio\ 2000; Ruiz-Lapuente et al.\ 2003).  In the SD channel, the
progenitor WD approaches the Chandrasekhar mass by accreting matter,
in a close binary system, from a companion star that is
thermonuclearly evolving (the nondegenerate component). The companion
could be a giant, a subgiant, or a main-sequence star.  In the DD
channel, the binary consists of two WDs which eventually merge,
thereby giving rise to the explosion; no bound object is left. In the
SD channel, on the other hand, the companion star should survive the
explosion and show distinguishing properties. The predictions of how
the companion star would look after the impact depend on the star's
physical properties before the explosion (Canal et al.\ 2001;
Livio\ 2000; Marietta et al.\ 2000; Podsiadlowski\ 2003; Pakmor et
al.\ 2011; Pan, Ricker, \& Taam\ 2012a,b; Shappee et al.\ 2013; Liu et
al.\ 2012, 2013).

To test the binary scenario, we observed and modeled stars within the
15\% inner radius of Tycho's SN 1572 (RL04).
%
%
Previous research (Ruiz-Lapuente\ 1997; Canal et al.\ 2001;
Ruiz-Lapuente et al.\ 2003) had pointed out that the most salient
feature of the surviving companion star should be peculiar velocities
with respect to the average motion of the other stars at the same
location in the Galaxy (mainly due to disruption of the binary),
detectable through radial-velocity and proper-motion measurements, and
perhaps also signs of the impact of the SN ejecta.
The latter can be twofold. First, mass should have been stripped from
the companion and thermal energy injected into it, possibly leading to
the expansion of the stellar envelope and making the star have a lower
surface gravity (Marietta et al.\ 2000; Podsiadlowski\ 2003).  Second,
depending on the interaction with the ejected material, the surface of
the star could be contaminated by the slowest-moving ejecta (made of
Fe and Ni isotopes).  Determination of the metallicity is also needed
in order to exclude the star belonging to the halo or thick disk. The
observations in RL04, therefore, were designed along these lines.

\begin{figure*}
\begin{center}
\includegraphics[width=178mm]{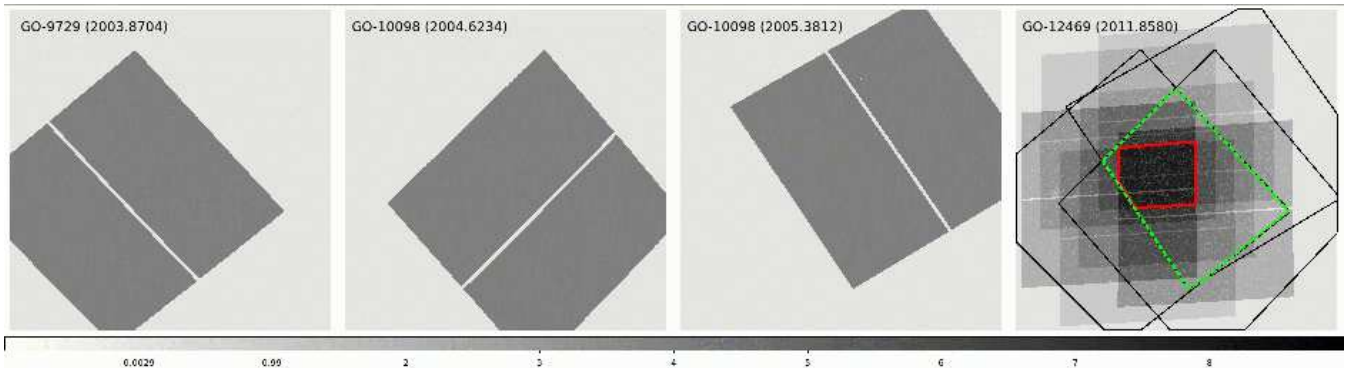}
\caption{
Depth-of-coverage map for the ACS/WFC and WFC3/UVIS epochs used in
this work.  Clearly, the WFC3/UVIS GO-12469 epoch (far-right panel) is
the most dithered one, and it provides the best overlap with the other
epochs.  Solid lines show the regions of interest of previous epochs,
while a green dashed line delimits the region for which four epochs
are available.  A subregion, marked in red, indicates where we have
the maximum number of deep exposures in all 4 epochs; sources within
it have the best PM measurements.
\label{deep}
}
\end{center}
\end{figure*}

%
The search for the binary companion of Tycho's SN 1572 has produced a
likely candidate: a G-type subgiant star labelled Tycho-G (RL04,
GH09). The star is relatively close to the centre of the supernova
remnant (SNR), its distance is compatible with being inside the SNR,
it has significantly higher radial velocity and proper motion (PM)
than stars at the same location in the Galaxy, and it shows signs of
pollution from the SN ejecta. The radial velocity, $v_r = -80 \pm 0.5$
km s$^{-1}$ in the Local Standard of Rest, is about 2\,$\sigma$ above
the average for stars at the distance of the SNR. Especially
significant, however, is the PM measured with the \textit{Hubble Space
  Telescope} (\hst\/) in programs GO-9729 (Cycle 12) and GO-10098
(Cycle 13): $\mu_l = -2.6 \pm 1.3$ mas yr$^{-1}$, $\mu_b = -6.1 \pm
1.3$ mas yr$^{-1}$.  At a distance of 3 kpc, this gives a tangential
velocity of $94 \pm 27$ km s$^{-1}$, but the disk velocity dispersion
is much smaller. It seems very difficult, therefore, to account for
the observed PM, except through some sort of binary interaction.

This point was later challenged by Kerzendorf et al.\ (2009, hereafter
K09). From comparison of a photographic plate taken in 1970 with the
Palomar 5-m telescope and a CCD image taken in 2004 (used in RL04)
with the 2.5-m Isaac Newton Telescope, K09 concluded that no
significant PM was detected: $\mu_l = -1.6 \pm 2.1$ mas yr$^{-1}$,
$\mu_b = -2.7 \pm 1.6$ mas yr$^{-1}$. However, the measurements prior
to the present work (RL04), also based on \hst\ images, again gave a
significant PM: $\mu_{\alpha \cos{\delta}} = -2.3 \pm 2.8$ mas
yr$^{-1}$, and $\mu_{\delta} = -4.8 \pm 2.8$ mas yr$^{-1}$ (maximal
external errors using the three epochs 2003, 2004, and 2005).  More
recently, a PM of about 5 mas yr$^{-1}$ was also obtained using the
same publicly available Cycle 16 \hst\ data
(Kerzendorf et al.\ 2013, hereafter K13), but the error bars
remain quite large.

Any improvement in the PM measurements of the stars significantly
increases information on the pre-explosion binary system. The peculiar
velocity of a surviving companion would correspond largely to its
orbital velocity just before the explosion, since the kick due to the
impact from the ejecta is a secondary effect (Marietta et al.\ 2000).
From the mass of the companion (the WD being near the Chandrasekhar
mass when it explodes), the orbit prior to the explosion can be
reconstructed, if we assume that both the radial and tangential
velocities of the centre of mass of the binary were typical for the
distance and Galactic latitude of the SNR.

With the present work we aim to refine the evolutionary path to the
explosion of SN 1572 by reducing by an order of magnitude the
uncertainties in the PMs of the stars. This improvement is now
possible because almost eight years have passed between the ACS/WFC
2003 images and the new ones (taken in late 2011, during \hst\ Cycle
19).

After the publication of RL04, there were claims that the proposed
companion of SN 1572, Tycho-G, was a giant star, with
log($g$/cm-s$^{-2}$) = $1.9 \pm 0.4$ dex and at a distance $\sim 10$
kpc (Schmidt et al.\ 2007). Keck HIRES spectra taken in 2006 did show
that Tycho-G is, in fact, rather similar to the Sun, as was already
discussed by RL04. The analysis of these new observations (GH09)
confirmed that it is a G-type subgiant, with $T_{\rm eff} = 5900 \pm
100$ K, log($g$/cm-s$^{2}$) = $3.85 \pm 0.30$ dex, and ${\rm [Fe/H]} =
-0.05 \pm 0.09$.  It is thus neither a halo star nor a typical
thick-disk star. Its distance is entirely compatible with that of the
SNR (around 3 kpc).  Moreover, a chemical abundance analysis showed a
clear excess of Ni (and, to a less significant extent, of Co), which
could point to possible contamination of the star's surface layers by
the SN ejecta (GH09).

Additional support for the single-degenerate origin of Tycho's SN has
been obtained recently from X-ray observations (Lu et al.\ 2011), and
the characteristics of the binary have been partially reconstructed
based on the results of RL04.  

Concerning the PM issue, very recently K13 have also analysed our now
publicly available GO-9729 and GO-10098 {\it HST} images and derived
values in agreement with previous results by RL04.
In particular, they confirm the high velocity of Tycho-G,
perpendicular to the Galactic plane.  In the present work, in which we
use {\it HST} observations spanning nearly 8 years, a similar result
is obtained, but with an almost tenfold precision.

\begin{table}
\caption{F555W data used in this work. }
\center
\begin{tabular}{ccc}
\hline
Epoch (date) / camera & $N_{\rm exp}$$\times$exp-time (s) & Dataset \\
\hline
                &                   &       \\
  2003.8704     & $4 \times 360$    & GO-9729  \\
 13 Nov.\ 2003  & $3 \times 10$      & GO-9729  \\
 ACS/WFC        & $3 \times 0.5079$  & GO-9729  \\
                &                   &       \\
  2004.6234 & $4 \times 360$     & GO-10098 \\
 15 Aug.\ 2004  & $3 \times 10$      & GO-10098 \\
 ACS/WFC        & $3 \times 0.5079$  & GO-10098 \\
                &                   &       \\
  2005.3812     & $4 \times 360$     & GO-10098 \\
 19 May   2005  & $3 \times 10$     & GO-10098 \\
 ACS/WFC        & $3 \times 0.5079$  & GO-10098 \\
                &                   &       \\
  2011.8580     & $1 \times 373$     & GO-12469 \\
  9 Nov.\ 2011  & $5 \times 372$     & GO-12469 \\
 WFC3/UVIS      & $2 \times 21$      & GO-12469 \\
                &                   &       \\
\hline
\end{tabular}
\label{tab1}
\end{table} 

Other stars close to the centre of Tycho's SNR are occasionally
suggested as possible companions, such as Tycho-E (Ihara et
al.\ 2007), which is, in fact, a double-lined binary (see GH09), or
Tycho-B (K13) because of its high rotational velocity, which is, in
fact, completely normal for its type: an A8--A9 main-sequence star
(Abt \& Morrell\ 1993, 1995).  More complete knowledge of the PMs of
all stars in the field, in addition to Tycho-G, was thus crucial (see
Sec.~9).

Here we first present and discuss the new astrometric work (Sec.~2 and 3).  
In Section~4 we redetermine the Ni abundance of Tycho-G, using the
same HIRES spectrum as in GH09 but a new procedure, and we compare it
with the Galactic trend, also defined by new high-quality data. We
reconstruct the binary orbit of the star (Sec.~5), assuming Tycho-G to
be the surviving SN companion, and from that we deduce its radius at
the time of the explosion, when it was filling its Roche lobe. In
Section~6, we discuss the kinematics of Tycho-G and its
significance. We evaluate the probability of picking at random a star
that combines all of its characteristics (Sec.~6). Sections~7 and 8
briefly discuss the uncertainties in the exact site of the SN
explosion and the luminosity evolution of a SN companion after the
impact of the SN ejecta. In Section~9, the data on the other 23 stars
with $V < 22$ mag and within $42''$ of the centroid of the SNR X-ray
emission are presented and discussed. We summarize our conclusions in
Section~10.

%
\section{Observations and Data Reduction}
\label{data}
%

The observations used to derive the PM measurements presented here
come from \hst\/ programs GO-9729, GO-10098, and GO-12469 (PI
Ruiz-Lapuente), which span $\sim 8$ yr.  Data were collected with both
the Wide Field Channel (WFC) of the Advanced Camera for Surveys (ACS),
and the Ultraviolet-Visual (UVIS) channel of the Wide Field Camera~3
(WFC3).
All of these images were taken with the filter F555W, available for
both cameras. However, we note that the two total transmissions
(instrument+filter) are slightly different.
For precise astrometric measurements and a more accurate assessment of
the uncertainties, we took particular care to dither our new WFC3/UVIS
images properly, with both large whole-pixel and fractional-pixel
offsets (following the general recipes given by Anderson \& King
2000).  This was not the case during previous ACS/WFC epochs,
resulting in some limitations to our achieved accuracies.
Table 1 lists both the ACS/WFC and WFC3/UVIS observations, while
Fig.~\ref{deep} shows the depth-of-coverage map for each of the four
epochs, after being transformed into the same reference system.

\begin{figure*}
\begin{center}
\includegraphics[width=178mm]{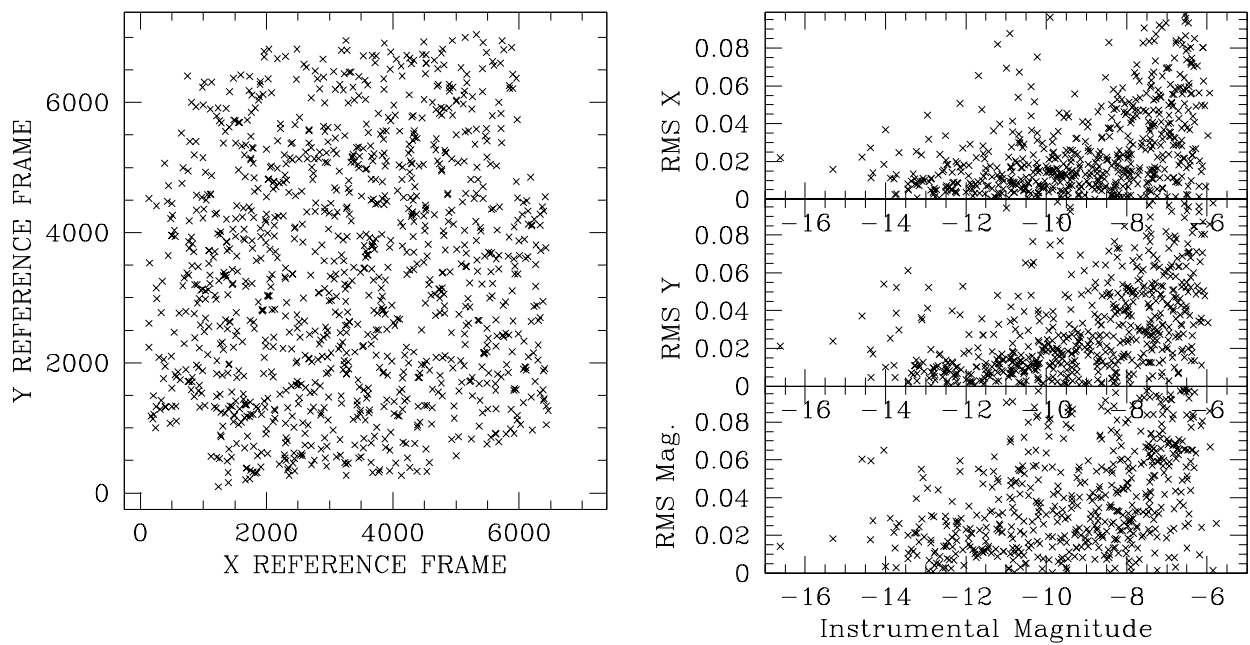}
\caption{
\textit{Left:\/} Spatial distribution of the reference frame built
from WFC3/UVIS data.
\textit{Right:\/} The RMS in $X$-coordinate position \textit{(top)},
$Y$-coordinate position \textit{(middle)}, and magnitude
\textit{(bottom)} as measured in single images (from 2, up to 8).
These RMS estimates have different noise, depending on the available
number of data points and the brightness of the sources. All positions
are in units of WFC3/UVIS pixels.
\label{rf}}
\end{center}
\end{figure*}

\begin{figure*}
\begin{center}
\includegraphics[width=178mm]{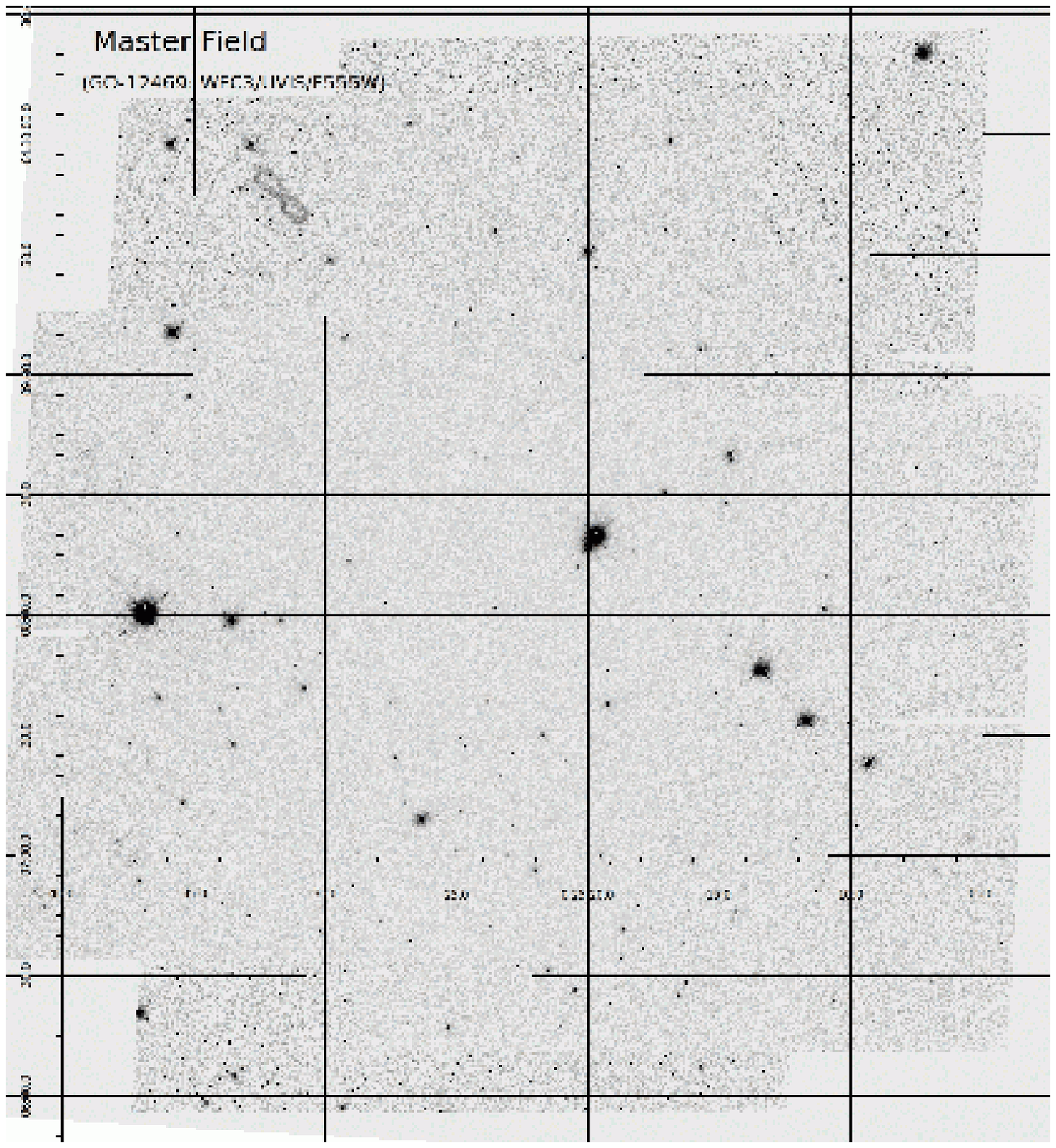}
\caption{
Stack of the deep WFC3/UVIS/F555W {\sf \_flt} images from GO-12469 
transformed into the reference frame.
\label{stack}            
}
\end{center}
\end{figure*}

\subsection{Correction for Imperfect CTE}
%

CCD detectors in the harsh radiation environment of space suffer
degradation owing to the impact of energetic particles, which displace
silicon atoms and create defects.  These defects temporarily trap
electrons, resulting in charge transfer efficiency (CTE) losses and in
trailing of the sources (because electrons are released at a later
time).  These effects have a major impact on astrometric projects
(Anderson \& Bedin 2010).

In this work, every single ACS/WFC image employed was treated with the
pixel-based correction for imperfect CTE developed by Anderson \&
Bedin (2010).  The algorithm is now improved\footnote{
The pixel-based CTE correction scheme based on the work of Anderson \&
Bedin (2010) has been modified to include the time- and
temperature-dependences of CTE losses (Ubeda \& Anderson 2012).  An
improved correction at low signal and background levels has also been
incorporated, as well as a correction for column-to-column variations.
} and directly included in the ACS pipeline. Standard calibrated
products are produced both with (\_{\sf flc} exposures) and without
(\_{\sf flt} images) this correction.
This correction proved to be effective in restoring fluxes, positions,
and the shape of sources, and reducing systematic effects of imperfect
CTE even in the worst case of extremely low backgrounds (Anderson \&
Bedin 2010; Ubeda \& Anderson 2012).
The WFC3 team is still working to develop and calibrate a similar
capability for UVIS; therefore, no CTE mitigation was operated on
WFC3/UVIS images in this work.  However, the relative ``youth'' of
UVIS's CCDs and the large-dither strategy of our new observations
provide a good handle on the astrometric biases resulting from
imperfect CTE, as well as on other systematic sources of errors.

\subsection{Fluxes and Positions in the Individual Images}
%
We measured positions and fluxes for every star in every \_{\sf flc}
ACS/WFC exposure, using a library of spatially variable effective
point-spread functions (PSFs) and the software programs documented by
Anderson \& King (2006).  Unfortunately, due to the sparseness of the
Tycho-G field, it was not possible to perturb the PSFs to account for
small focus variations.  (We will see in Sec.~\ref{bc} how we have
mitigated this and other systematic biases in our astrometry.)
As in Bedin et al.\ (2003, 2006), we used the best available average
distortion corrections (Anderson 2002, 2005) to correct the raw
positions and fluxes of sources that we had measured within each
individual ACS/WFC {\_flc} exposure.

Positions and fluxes of sources in each WFC3/UVIS \_{\sf flt} image
were obtained with software that is adapted from the program 
{\sf img2xym\_WFI} (Anderson et al.\ 2006). Astrometry and photometry
were then corrected for pixel area and geometric distortion using the
best available average distortion corrections and library PSFs
(Bellini \& Bedin 2009; Bellini et al.\ 2011).

\section{Astrometry and Proper Motions}
\label{aapms}
%

To derive proper motions, we essentially follow the detailed
procedures given by Anderson \& van der Marel (2010, hereafter Av10)
to measure the internal motions of the stars in the core of the
populous globular cluster $\omega$ Centauri. However, we need to take
into account that the Tycho-G field has major differences with respect
to the field studied by Av10.

Our astrometric measurements, like those of Av10, are relative to a
group of objects in the observed field. As a reference, Av10 used
cluster members; there are hundreds of stars per square arcminute
having high signal-to-noise ratios (S/N), all at a common distance,
and sharing a common PM to within a few 0.1 mas yr$^{-1}$ (the
internal velocity dispersion of $\omega$ Cen is less than 20 km
s$^{-1}$).

By contrast, our Tycho-G field is sparsely populated, with only a few
objects per square arcminute. Furthermore, the reference stars are at
different distances and have different PMs.  We also note that no
obvious and suitable background object is present in our field, as a
result of the high extinction at low Galactic latitudes ($b \approx
1.5^\circ$).
In our Tycho-G field, the uncertainties in our PMs are dominated by
the PM dispersions of the stars with respect to which we measure the
positions at the different epochs, even selecting stars that moved the
least with respect to each other.
Nonetheless, our long temporal baseline of $\sim 8$ yr improves the
final precision to levels comparable to those reached by Av10.

   \begin{figure}
   \begin{center}
   \includegraphics[width=85mm]{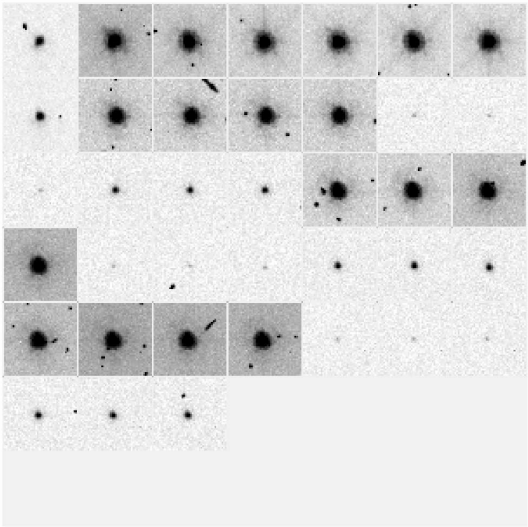}
   \caption{
From left to right, and from top to bottom, a mosaic of the $40 \times
40$ pixels around Tycho-G in each of the 38 images available to us
(see Table~\ref{tab1}).  The first eight are from the UVIS epoch, and
the next 30 from ACS (specifically, 10 collected in 2003, then 10 from
2004, and the last 10 from 2005).
   \label{imagette}
}
   \end{center}
   \end{figure}

\subsection{The Reference Frame}
%

The first important step is to build a distortion-free
\textit{reference frame} using all of the images taken within a chosen
epoch (the \textit{reference epoch}), with respect to which we will
later perform \textit{all} of the relative measurements.
For this task we selected our recent WFC3/UVIS epoch, as this epoch is
the one with the largest number of deep observations, optimal
dithering, and covering the largest fraction of the other previous
epochs (see Fig.~\ref{deep}). In addition, because of its better pixel
sampling, WFC3/UVIS intrinsically provides better imaging astrometry
($\sim 0.3$ mas; Bellini et al.\ 2011) compared with ACS/WFC ($\sim
0.5$ mas; Anderson \& King 2006).

To build the reference frame, we began by identifying bright,
isolated, and unsaturated sources measured within each WFC3/UVIS
\_{\sf flt} image of program GO-12469.
Among these, we then selected those having a stellar profile.  To
measure the stellarity of objects, we use the ``quality-fit'' ({\sf
  q-fit}) parameter described by Anderson et al.\ (2008), which
essentially quantifies how close the distribution of the observed
pixel values resembles the local PSF model.
This selection allowed us to immediately reject most cosmic ray (CR)
hits, warm pixels, artifacts, and potentially also extended nonstellar
objects.  With caution, this enabled us to use those portions of the
field observed only once in the reference epoch (i.e., those regions
for which there is only one exposure in the reference epoch).

We initially took as a reference the distortion-corrected positions
measured in the deep image \textsf{iboy01mhq}, which is at the centre
of the WFC3/UVIS dither pattern, and linearly transformed the star
positions (distortion-corrected) from all of the other images into
that frame. We then determined an average position for each star that
was found in at least five exposures.
In doing this, we used the most general linear transformations (6
parameters).

To improve this reference frame, we found a linear transformation from
each exposure into the new frame, based on the positions of common
stars.
For each star in the reference frame, we thus had between five and
eight estimates for its position (depending on how many images
overlapped at that point), and we averaged these positions together to
improve the reference frame.
After a few such iterations, the root-mean square (RMS) of these
multiple (and dithered) observations was less than 0.01 pixels for the
brightest stars (with 5 or more observations).

To extend the reference frame where there are fewer than five images,
we used the same transformations obtained above to include the objects
with faint magnitudes and a stellar profile, and stars measured only
in four, three, two, and one image.
The resulting reference frame contains 1148 objects (a few of which
are artifacts or remaining CRs). Its spatial distribution, together
with the RMS in magnitude and in $X$ and $Y$ positions, are shown in
Fig.~\ref{rf}.

\subsection{Absolute Astrometry}
%

As extensively discussed in Sec.~\ref{bc}, the adopted geometric
distortion correction is just an average solution, and from frame to
frame there are sizable changes. This is particularly true for the
linear terms, which contain the largest portion of the variations.
So far, in deriving the reference frame, we used 6-parameter linear
transformations to register the distortion-corrected positions
measured in each frame to the distortion-corrected positions measured
in the reference image \textsf{iboy01mhq}; the linear-term variations
with respect to the reference image were completely absorbed by the 6
parameters.
Therefore, the astrometric zero points, plate scales, orientations, and
skew terms of our adopted reference frames are still not calibrated to
an absolute reference system.

We used sources in common between our reference frame and the Two
Micron All Sky Survey (2MASS, Skrutskie et al.\ 2006) to determine the
unconstrained linear terms.
Within our Tycho-G region we found $\sim 80$ 2MASS point sources in
common with our reference frame which we used to calibrate our linear
terms.  The constrained linear terms enabled absolute astrometry
accurate to $\sim 0.2''$.
As previously mentioned, the relative positions of stars are much more
accurate than their absolute zero points.
The nonlinear part of the ACS/WFC and WFC3/UVIS distortion solutions
is accurate to $\sim 0.01$ original-size WFC3/UVIS pixel ($\sim
0.4$--0.5 mas) in a global sense (Anderson \& King 2006; Bellini et
al.\ 2011), roughly the random positioning accuracy with which we can
measure a bright star in a single exposure. Recently, it has been
discovered that the linear terms of the ACS/WFC distortion solution
have been changing slowly over time (Anderson 2007). Even if this were
the case also for WFC3/UVIS, we note that in our procedure the linear
terms are constrained by the 2MASS catalog.
The absolute coordinates are referred to equinox J2000.0, with
positions given at the reference epoch, 2011.858.

\subsection{Image Stack}
%

With the transformations from the coordinates of each image into the
reference frame it becomes possible to create a stacked image of the
field within each epoch.
The stack provides a representation of the astronomical scene that
enables us to independently check the region around each source at
each epoch.
The stacked images are 15,000 $\times$ 15,000 super-sampled pixels (by
a factor 2; i.e., 20 mas pixel$^{-1}$), and corresponding to $\sim 5'
\times 5'$.  The stack for the WFC3/UVIS 2011 epoch is shown in
Fig.~\ref{stack}.
We have included in the header of the image, as World Coordinate
System (WCS) keywords, our absolute astrometric solution based on the
2MASS point-source catalog, as described in the previous section.
As part of the material provided in this paper, we give the
astrometrised stack-image for the reference epoch electronically
online.

In Fig.~\ref{imagette}, we show a mosaic of the $40 \times 40$
pixel field for all 38 images, centred on Tycho-G.

   \begin{figure*}
   \begin{center}
   \includegraphics[width=168mm]{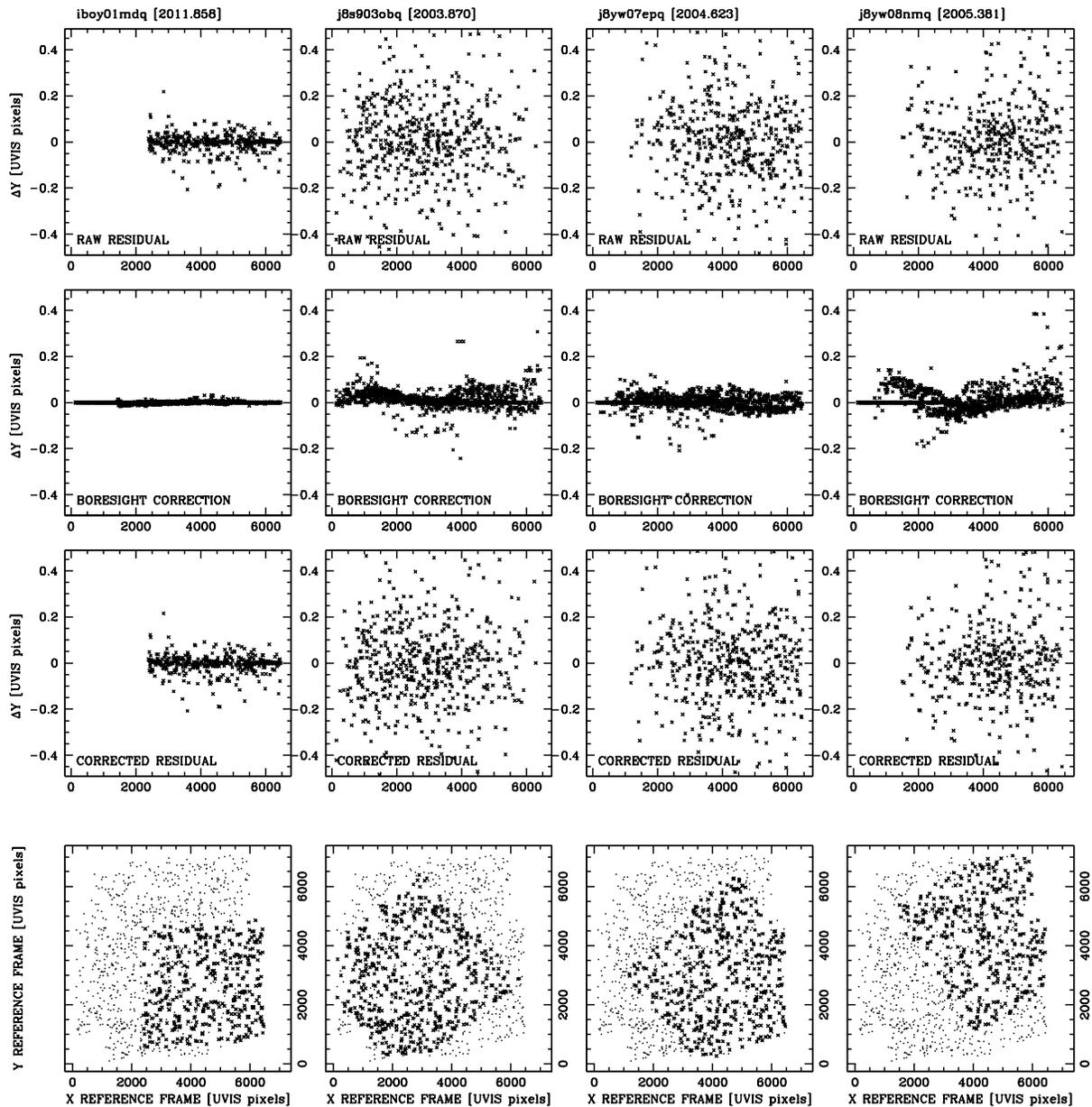}
   \caption{
   Each column of panels refers to the first deep image of each epoch. 
   \textit{Top panels:} As a function of the $X$-coordinate in 
   the reference frame, we show the $Y$-residual (raw) between positions of
   stars measured in an individual image (distortion corrected and linearly 
   transformed into the reference frame), and its $Y$-position in the reference 
   frame.
   \textit{Second panels:} The local ``boresight'' correction 
   used for each star (see text) to remove residuals in the distortion 
   correction. 
   \textit{Third panels:} The residuals after making these local 
   adjustments. 
   Those in the left column (relative to image {\sf iboy01mdq}) 
   show only distortion errors (plus random errors) but have no PMs, 
   as the reference frame is based on images taken at that epoch.  
   For exposures taken at the other epochs (other columns), these
   residuals contain both distortion errors and actual PMs,
   and therefore show larger scatter than in the left column.
   Note that many of these point sources are faint stars that have 
   large random errors due to the low S/N. Also, these are 
   single-image measurements, which could be affected by CRs and other 
   artifacts. 
   \textit{Bottom panels:} The location of the image stars (crosses)
   with respect to the master-frame stars (thin dots).
   \label{bore}
   }
   \end{center}
   \end{figure*}

\subsection{Boresight Correction}
\label{bc}
%

Positional imaging astrometry in \hst\/ images is so precise that it
allows us to appreciate small variations from frame to frame, even if
collected consecutively within a same orbit.
For example, the velocity of \hst\/ around the Earth ($\pm 7$ km
s$^{-1}$) causes light aberration, inducing plate-scale variations up
to 5 parts per 100,000 (Cox \& Gilliland 2003) that can be measured
clearly (Anderson et al.\ 2007; Bellini et al.\ 2011).

There are also less predictable time-dependent changes caused by the
temporal variation of the \hst\/ focus. These focus variations are
correlated with thermal variations induced by the angles between the
Sun and the telescope tube (the so-called ``breathing''); they result
in changes in the geometric distortion and in position-dependent PSF
variations, which in turn are the result of a complicated interplay
between aberrations and charge diffusion across the detector (Jee et
al.\ 2007).
Therefore, the astrometric distortion solutions provided by Anderson
\& King (2006) for ACS/WFC and by Bellini et al.\ (2009, 2011) for
WFC3/UVIS should be intended only as \textit{average} geometric
distortion solutions.  The same considerations are also valid for the
adopted library (and thus \textit{average}) PSFs.
Our estimated positions could thus change appreciably from frame to
frame when using library PSFs and average distortion solutions.

To account for these small but sizable effects, Av10 introduced local
adjustments to the measured positions, which they call ``boresight''
corrections.
These corrections are determined for each star in each frame as
follows. For each exposure that included the star, we calculate a
robust average offset between the globally transformed (into the
reference frame) positions of the neighbouring stars and the average
reference-frame positions of the same neighbouring stars.  This
average offset provides the correction to the transformed position of
the target star for that exposure.
Hence, the reference frame contributes only to the plate scale and
orientation (see Av10 for details).
For images collected in the same epoch as the reference frame, we
selected among neighbouring stars only those having positions within
0.05 WFC3/UVIS pixels from their positions in the master frame, but
within 1 WFC3/UVIS pixel for the other ACS/WFC epochs. (We must be
more generous from epoch to epoch, as the displacements also contain
the intrinsic motions of stars.)

In the case of Av10 these corrections were extremely local, over a few
tens of pixels, since the set of reference stars was very dense.
But owing to the sparseness of the Tycho-G field, our local set (used
for the boresight corrections) includes neighbouring stars up to 1000
UVIS pixels from the target stars (i.e., a $\sim 1/4$ of a WFC3/UVIS
field), thereby making these corrections not very ``local.''
Thus, our boresight correction will remove only residuals with this
spatial scale.
Also, the boresight correction is calculated only if at least 6
suitable objects are available within 1000 UVIS pixels from the
target. Typically we had $\sim 40$ objects, but for short exposures
the paucity of stars forced us to accept as few as 6.

The top panels in Fig.~\ref{bore} show the ``plain'' (or raw)
residuals between positions measured in individual frames and the
reference frame. The corresponding boresight corrections are given in
the second panels from the top, while the ``boresight-corrected''
residuals are shown in the third panels.  The bottom panels display
the spatial distributions of the sources with respect to the reference
frame.
We note that these boresight corrections are small within the
reference epoch, of order $\sim 0.01$ WFC3/UVIS pixels ($\sim
0.4$ mas), but they can be as large as 0.1 pixels ($\sim 4$ mas) from
one epoch to another.

Here we must make an important consideration concerning the achievable
accuracy of our PMs ($\mu$), which are described in the next section.
The accuracies we can hope to achieve ($\overline{\sigma_{1D\mu}}$)
are ultimately set by \textit{both} the PM dispersion
($\sigma_{1D\mu}$) and the number of neighbouring stars ($N_{\rm
  used}$) used to derive the boresight corrections. These accuracies
can be formulated in the relation [for each of the two 1D PM
  components, ($\mu_{\alpha\cos{\delta}}$,$\mu_{\delta}$)] as
\begin{equation}
\overline{\sigma_{1D\mu}} = \frac{\sigma_{1D\mu}}{\sqrt{N_{\rm used}-1}}.
\label{eq1}
\end{equation}
This statistical ``kinematic noise'' of the network of stars, whose
positions are not fixed with respect to each other, dominates our
uncertainties.  We will further discuss this issue in
Sec.~\ref{error}, where we also explicitly quantify the values of
$\overline{\sigma_{1D\mu}}$ for Tycho-G.

   \begin{figure}
   \begin{center}
   \includegraphics[width=85mm]{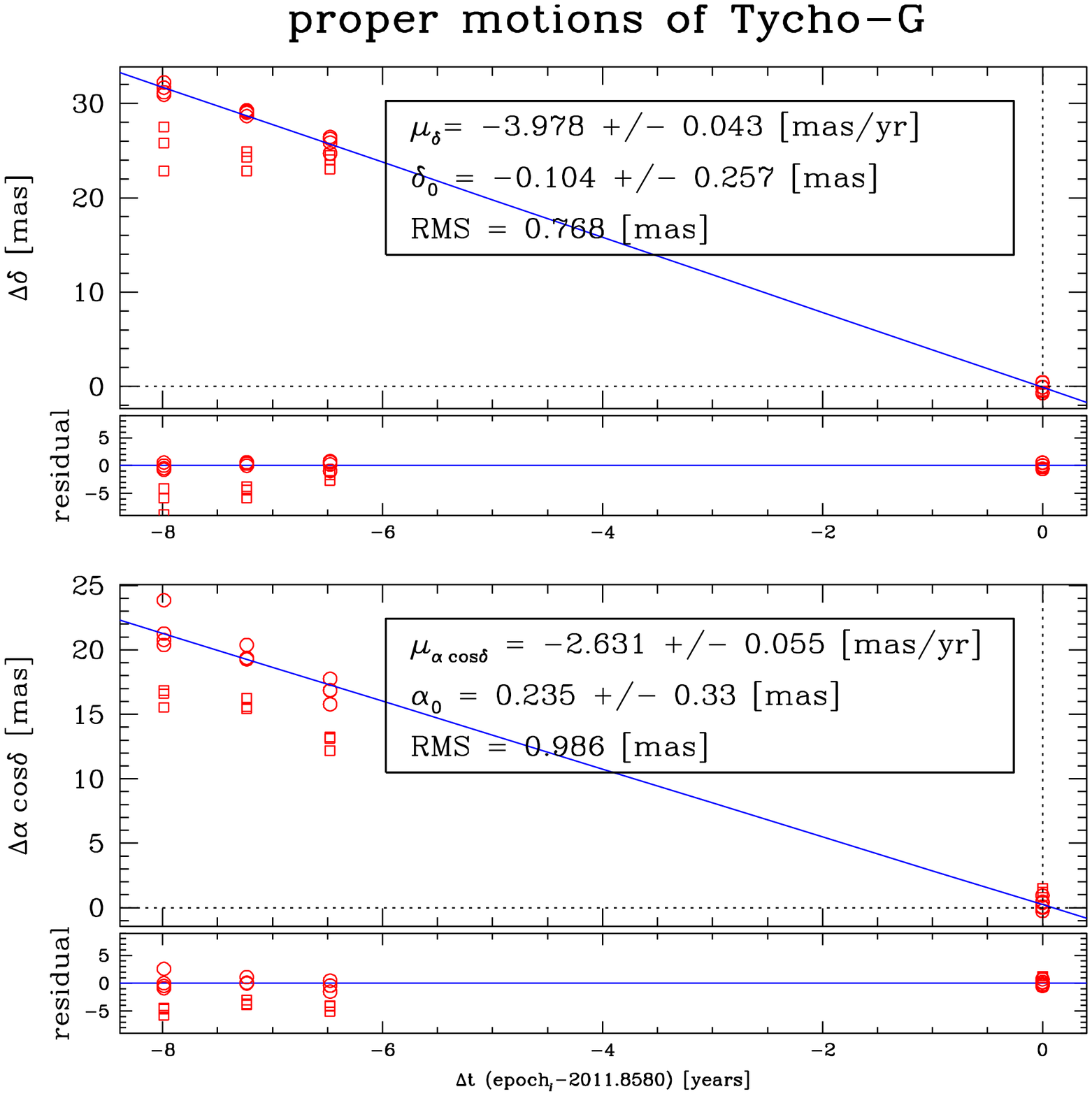}
   \caption{
Multi-epoch fit of the PMs (and residuals) for Tycho-G (bottom half
for the motion in $\alpha\cos{\delta}$, and top half for the motion in
$\delta$).
In each half, the upper panel shows the displacement (in mas) for the
four epochs as a function of time from the reference epoch (in years).
The blue line is the weighted best fit to the data points (circles
from deep exposures and squares from short exposures), and its slope
is the PM.
The quoted errors are formal values of the fit (see text for a
discussion of the actual uncertainties).
The lower panel shows the residuals to each fit. 
   \label{figTyG}
            }
   \end{center}
   \end{figure}

   \begin{figure}
   \begin{center}
   \includegraphics[width=85mm]{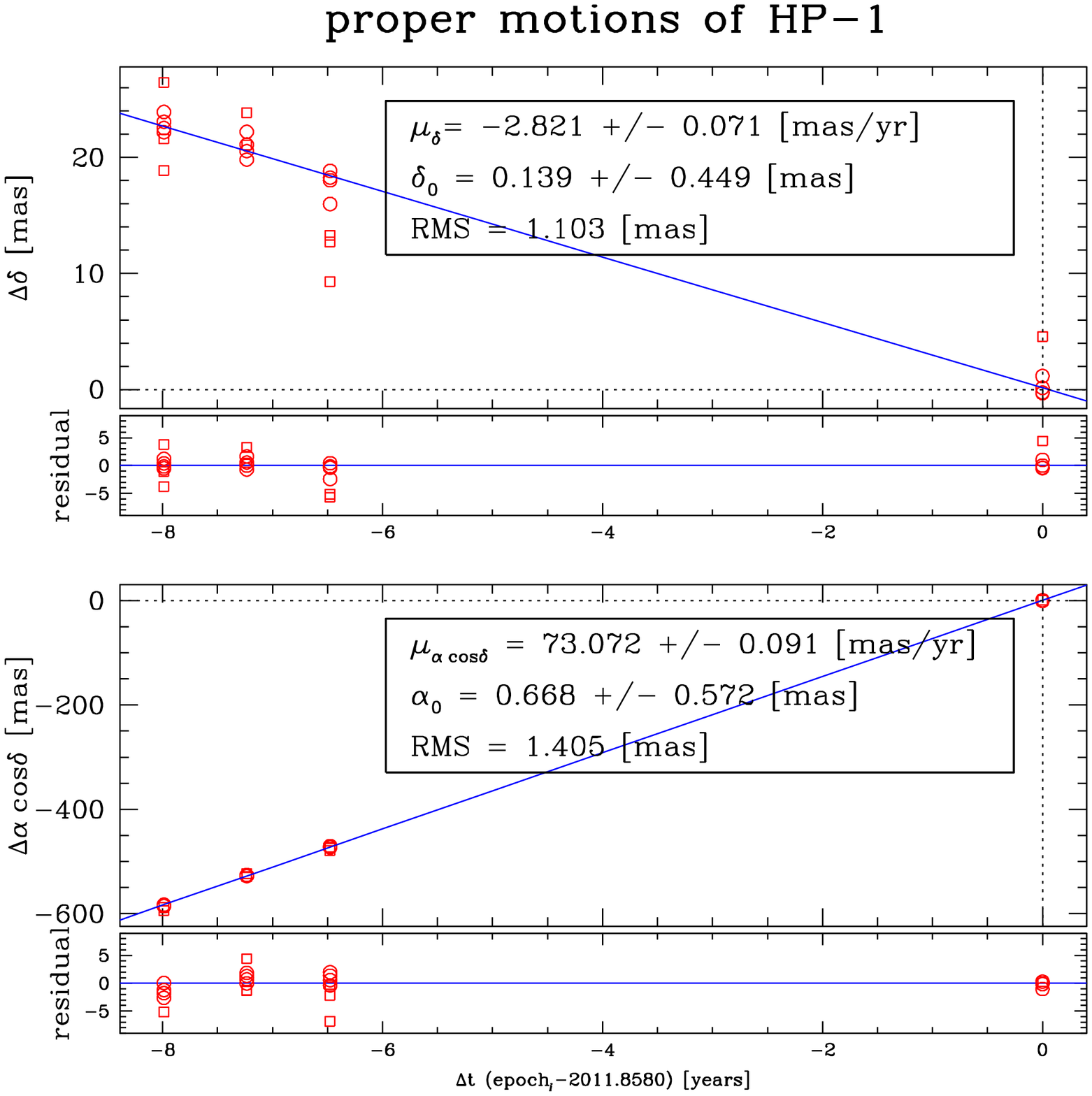}
   \caption{
Same as in Fig. \ref{figTyG} but for HP-1. 
   \label{figHP1}
            }
   \end{center}
   \end{figure}

\subsection{Proper-Motion Measures}
\label{pms}
%

In Fig.~\ref{figTyG} we summarise our multi-epoch fit to the motion of
Tycho-G.  The bottom and top halves refer to the $\alpha\cos{\delta}$
and $\delta$ motions, respectively.  In each half, the top panel shows
the boresight-corrected displacements as measured at the various
epochs (i.e., residuals as those in the third-panel from top in
Fig.~\ref{bore}).  The displacements are with respect to the reference
frame (in mas units), and the time (in years) is relative to our
reference epoch (2011.8580).
The data points from deep exposures are indicated with (red) open
circles, while those from short exposures with (red) open squares.
Data points from saturated images are not used.
The blue lines are our weighted best {\it linear} fit to the data
points, where for weights we adopted the raw fluxes measured in the
individual images.
In this way, when both short and long exposures are available (as is
the case for Tycho-G), data points coming from short exposures receive
negligible weights. Indeed, short exposures are the most affected by
CTE, and by statistical limitations related to the paucity of
reference stars having good S/N.\footnote{
The disagreement between the motions inferred from short and long
exposures for the case of Tycho-G should be used as a warning for
potential systematics in the motions of other objects based only on
short exposures.
}
Therefore, the slopes of the blue lines are our estimates for the PMs
of Tycho-G.  These are labeled in the figure as
$\mu_{\alpha\cos{\delta}}$ and $\mu_\delta$ with their formal errors.
The average of the displacements should be null at the reference epoch. 

Note that in fitting the data, we let the positions at the reference
epoch float freely; consequently, $\delta_0$ and $\alpha_0$ indicate the
corrections to be applied to our reference-frame positions at epoch
2011.858. However, these small corrections make sense only in relative
terms, the uncertainties in the absolute positions being dominated by
the uncertainties in the 2MASS absolute zero points.
The lower panels in each half of Fig.~\ref{figTyG} show the residuals to 
the fit, and the RMS of these residuals is labelled.

We attempted an analogous fit to all of the sources in the master frame. 
As an example of these, Fig.~\ref{figHP1} shows our derived PMs for
another object of interest in the field, the high-PM source HP-1
(K09).
It is interesting to note that spectra of HP-1 suggest it is a late-M
star, at a maximum distance of 500 pc (i.e., neglecting interstellar
extinction); therefore, the geometric parallax should show a deviation
from the linear fit of at least 2 mas.

It is worth mentioning that our derived PMs are consistent with those
found by K13 (for the objects in common), although with much higher
precision given our increased time baseline.


In the left panel of Fig.~\ref{VPD} we show the vector-point diagram
(VPD) for the 872 objects (out of 1148) in the reference frame for
which it was possible to derive relative proper motions. The adopted
zero point of the motion is indicated by dotted lines.
For 222 of them (indicated in black with their error ellipses), four
epochs were employed, while for 296 (indicated in blue), just three
epochs.  Objects having uncertainties larger than 0.25 mas yr$^{-1}$
are indicated with crosses. For reference, a red circle at (75,45) mas
yr$^{-1}$ indicates the 0.25 mas yr$^{-1}$ uncertainties.
For 354 objects there are only two epochs (magenta). 
Two arrows indicate the motions of Tycho-G (red) and HP-1 (green).  
A close-up view of the VPD around the zero motion is shown Fig.~\ref{VPDzoom}.
The right panel of Fig.~\ref{VPD} gives the spatial distribution of
sources in the reference frame, for objects with four epochs (black),
three epochs (blue), two epochs (magenta), and only one epoch
(grey). Tycho-G is highlighted in red and HP-1 in green.

\subsubsection{Uncertainties} 
\label{error}

The formal uncertainties provided by the weighted linear fit are
probably much too optimistic, because our internal estimates of the
errors within a given epoch (and based on multiple observations) could
be severely underestimated.  This is almost surely the case for the
nondithered ACS/WFC epochs, where a portion of the systematic errors
might cancel out and do not appear in the estimated internal errors.
%

   \begin{figure*}
   \begin{center}
   \includegraphics[width=84mm]{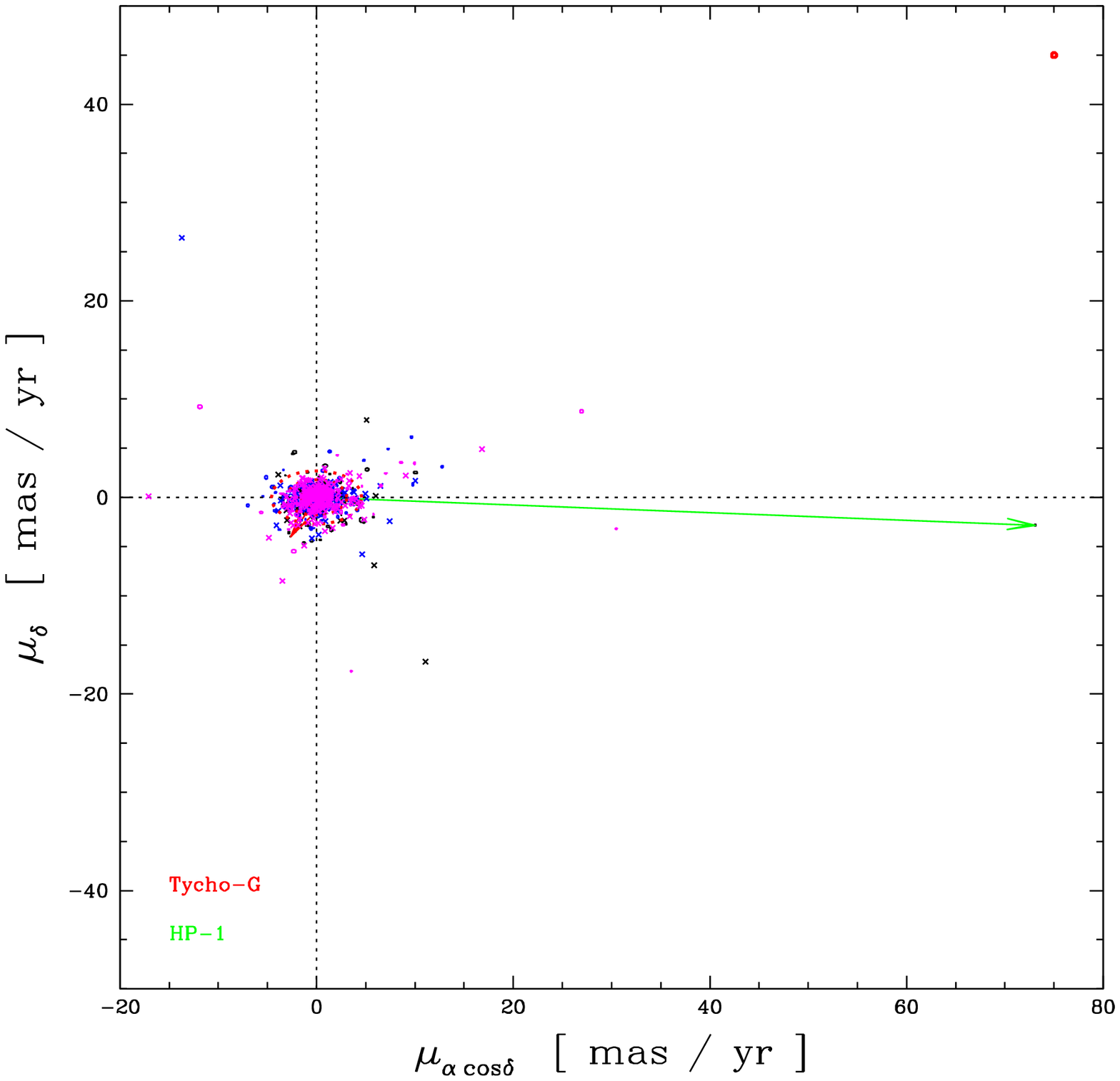}
   \includegraphics[width=84mm]{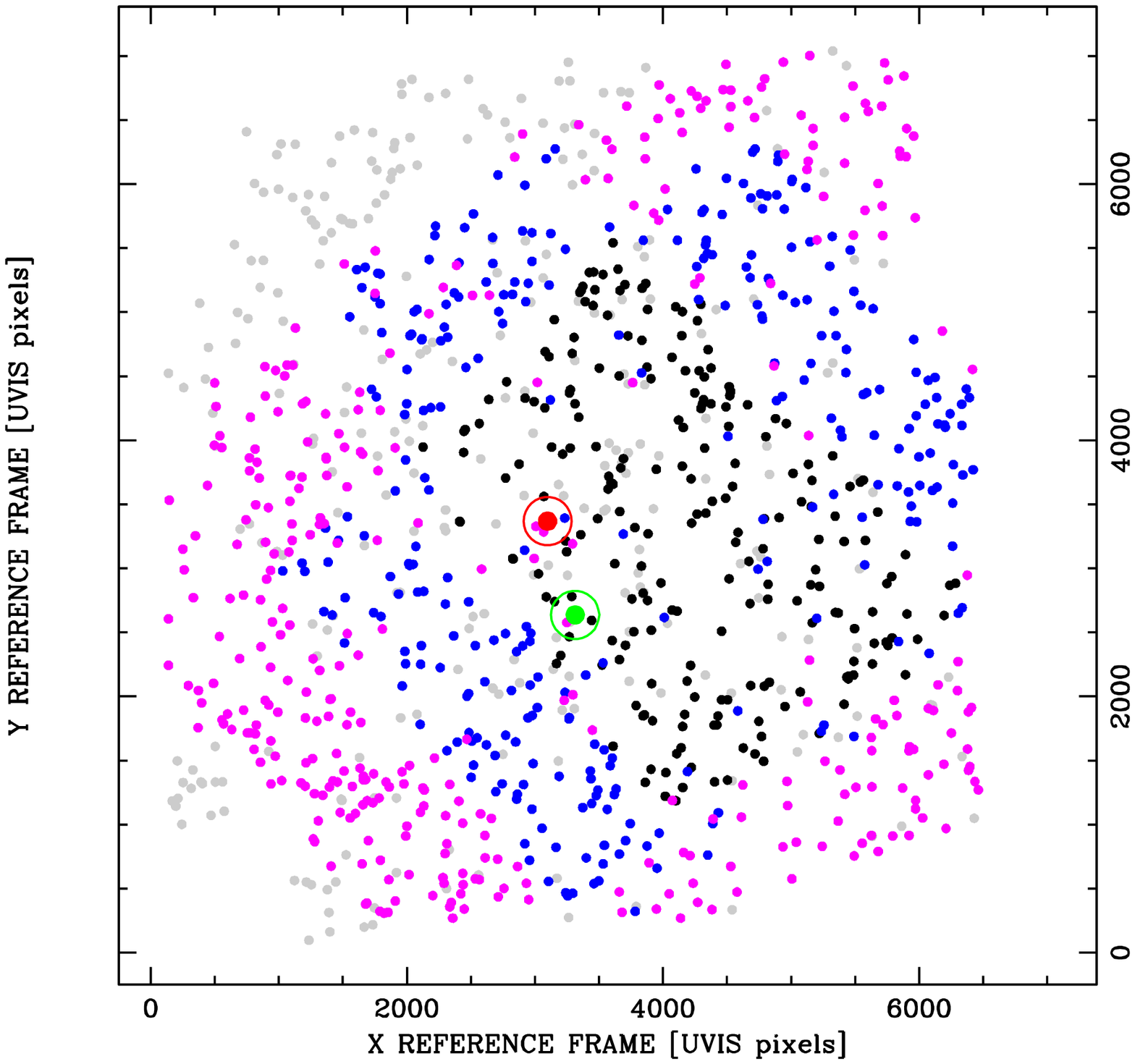}
   \caption{
\textit{Left:\/} Vector-point diagram for the 872 objects (out of
1148) in the reference frame for which it was possible to derive
relative PMs.
For 222 of them (indicated in black with their error ellipses) four
epochs were employed, while for 296 (indicated in blue), three.
For clarity, those with uncertainties larger than 0.25 mas yr$^{-1}$
are indicated as crosses, and a red circle at position (75;45) mas
yr$^{-1}$ shows for reference the 0.25 mas yr$^{-1}$ error.
For 354 objects there are only 2 epochs (thin magenta points), and
therefore no external estimates of the uncertainties were available;
we simply combined in quadrature the RMS within each of the two
epochs.
Two arrows indicate the motion of Tycho-G (red) and HP-1 (green). 
The zero motion of the reference frame is indicated by dotted lines. 
A close-up view of the zero motion is shown Fig.~\ref{VPDzoom}. 
\textit{Right:\/} Spatial distribution of sources in the reference
frame having four epochs (black), three epochs (blue), two epochs
(magenta), and only one epoch (grey).  Tycho-G is highlighted in red
and HP-1 in green.
   \label{VPD}
}
   \end{center}
   \end{figure*}

But there is a more fundamental limitation.  As mentioned at the
beginning of Sec.~\ref{aapms}, and more clearly stated at the end of
Sec.~\ref{bc}, our PM accuracies are ultimately set by the PM
dispersions of the set of stars with respect to which we have measured
the relative positions --- that is, the neighbouring stars used for
the boresight corrections.
A more realistic estimate of the errors is provided by Eq.~\ref{eq1},
given at the end of Sec.~\ref{bc}, which requires the estimates of
$\sigma_{\mu_{\alpha\cos{\delta}}}$ and $\sigma_{\mu_{\delta}}$.

To estimate these for each coordinate, we begin by taking the
68.27$^{\rm th}$ percentile around the zero of the absolute value of
the PMs obtained in the previous section (and shown in Fig.~\ref{VPD})
as an initial guess of the PM dispersions ($\sigma_\mu$). We then
clipped at 3.5 times this value the objects with higher PMs, assuming
them to be outliers. We then redetermined the dispersion of the PMs
from the purged sample, again by taking the 68.27$^{\rm th}$
percentile as a robust estimate of the PM dispersion (assumed as a
first approximation to be Gaussian). We did this for both
$\alpha\cos{\delta}$ and $\delta$ (using all PMs based on 4, 3, and 2
epochs), obtaining $\sigma_{\mu_{\alpha\cos{\delta}}} = 1.09$ mas
yr$^{-1}$ and $\sigma_\delta = 0.62$ mas yr$^{-1}$.
The larger dispersion in $\mu_{\alpha\cos{\delta}}$ than in $\mu_\delta$ is
caused by the almost exact alignment between $\alpha$ and the Galactic Plane 
at the location of Tycho-G. 

Concerning the term $N_{\rm used}$ in Eq.~\ref{eq1}, we took the
average number of neighbouring stars used to compute the boresight
correction in the adopted frames: $N_{\rm used} = 39$.  Therefore, a
more realistic estimate for the expected uncertainties in our PMs is
$(\overline{\sigma_{\mu_{\alpha\cos{\delta}}}};\overline{\sigma_{\mu_{\delta}}})
=
(\sigma_{\mu_{\alpha\cos{\delta}}}/\sqrt{39-1};\sigma_{\mu_{\delta}}/\sqrt{39-1}))
= (0.18;0.10) \textrm{~ mas yr}^{-1}.$
Thus, our estimated relative proper motion for Tycho-G is 
$$(\mu_{\alpha\cos{\delta}};\mu_\delta)_{\rm J2000.0} = $$ 
$$(-2.63;-3.98) \pm (0.06;0.04)\pm (0.18;0.10) ~ \textrm{mas yr}^{-1},$$
where the first terms in the uncertainties are the formal errors,
which are negligible relative to the expected errors given by the
second term.
It is worth noting that these values for Tycho-G are consistent with
those recently obtained by K13, although at a much higher accuracy
than those of K13, thanks to our larger time baseline.

To reiterate, the dominant uncertainties take into account that all of
our measurements of positions and PMs of the target stars are relative
to the average motions of the bulk of the other stars in the field,
and specifically to the stars used to compute the boresight
corrections.
Note that even had we measured the relative positions of our sources
(within each epoch) with infinite precision, this dominant ``kinematic
component'' to the error budget would still remain, as our positions
are measured with respect to an ensemble of objects which are actually
\textit{not} fixed.

We conclude this section by noting that the residuals to the fit,
shown in the bottom panels of Figs.~\ref{figTyG} and \ref{figHP1}, are
caused by unaccounted sources of error, but in principle could also
reflect the geometric parallaxes (of course \textit{relative}
parallaxes and not absolute parallaxes, as for all the other
measurements in this work).  Although theoretically possible, we have
not attempted here to solve for two positions ($\alpha_0,\delta_0,$),
two proper motions ($\mu_{\alpha\cos{\delta}},\mu_\delta$), and the
parallax ($\pi$) with just 6 independent data points [(4--1) epochs
  $\times$ 2 coordinates].

\section{The Ni Abundance of Tycho-G Revisited}

The Ni abundance of Tycho-G was first determined as [Ni/Fe] $= 0.16
\pm 0.04$, using a high-resolution Keck-I/HIRES spectrum (GH09).
Recently, K13 reanalysed the same spectrum and derived [Ni/Fe] $= 0.07
\pm 0.04$.  They argue that this difference could be explained by
differences in equivalent width (EW) measurements of Ni lines which
may be related to continuum normalisation and/or local continuum
placement.
 
We have now revised the Ni abundance found by GH09 using automatic
tools able to search for and fit the continuum, and to measure the EWs
of Ni lines in the HIRES spectrum of Tycho-G. We use the code
\textsf{ARES} (Sousa et al.\ 2007) and the Ni line list from GH09,
which comes from the abundance analysis of Gilli et al.\ (2006).
Within the \textsf{ARES} code we choose the value 0.965 for the
``rejt'' parameter to take into account the relatively low S/N ($\sim
30$) of the HIRES spectrum. After running \textsf{ARES}, we also
filter the output line list, requiring the Gaussian FWHM of the output
fitted Ni lines to be in the range 0.08--0.25~\AA\ in order to avoid
wrong fits. The instrumental broadening of this spectrum is about
0.14~{\AA} at $\lambda = 6000$~{\AA}, but we allow this range because
of the relatively low S/N.

K13 estimate the average uncertainty in the EW determination to be
6.5~m{\AA}. We thus decided to remove all weak lines with EWs below
15~m{\AA}, to avoid uncertain measurements.  We also apply a very
restrictive 1.5\,$\sigma$ clipping in the Ni abundance values to discard
the possible outliers. The final result including 12 Ni lines gives
$A$(Ni) = log[$N$(Ni)/$N$(H)] + 12 = 6.28 ($\sigma = 0.11$,
$\Delta_\sigma = 0.03$), where $\sigma$ is the dispersion of the Ni
abundance from $N$ lines and $\Delta_\sigma= \sigma/\sqrt{N}$.

We now want to compare the Ni abundance of Tycho-G with high-quality
abundance data from stars observed with the HARPS spectrograph on the
3.6-m telescope at La Silla Observatory (Neves et al.\ 2009).

We cross-check our line list with the Ni line list of Neves et
al.\ (2009) and find 8 lines in common. The derived Ni abundance is
$A$(Ni) = 6.30 ($\sigma = 0.13$, $\Delta_\sigma = 0.05$, $N=8$)
according to this new line list.  We measured the EWs of these 8 Ni
lines in the Kurucz solar ATLAS (Kurucz et al.\ 1984) to derive the
solar Ni abundance and obtained $A$(Ni)$_\odot = 6.25 \pm 0.00$, which
we use to determine the [Ni/Fe] $\equiv$ log[$N$(Ni)/$N$(Fe)] $-$
log[$N$(Ni)/$N$(Fe)]$_\odot$ ratio. Therefore, taking into account the
metallicity of Tycho-G, [Fe/H] $ = -0.05$, our revised Ni abundance is
[Ni/Fe] $ = 0.10 \pm 0.05$.\footnote{
Note that the uncertainty (0.05) has slightly increased compared with
previous work by GH09 (0.04), mostly because of abundance dispersion
(since other sources of error cancel out when computing the abundance
ratio [Fe/Ni]).  Also, we now use updated oscillator strengths, and we
compare our derived abundance with a new high-quality set of
abundances.  The reference Galactic trend (Neves et al.\ 2009) appears
to be slightly higher than before (Gilli et al.\ 2006), probably due
to the different choice of Ni lines and different oscillator-strength
values.  
}
%
%
This revised Ni abundance is between the values given by GH09 
and K13, but is consistent within the error bars with both values. 
We think that this value may be more accurate than that of GH09, 
since it was obtained using automatic tools which may be 
more appropriate when dealing with low-S/N spectra. 
%

   \begin{figure}
   \begin{center}
   \includegraphics[width=85mm]{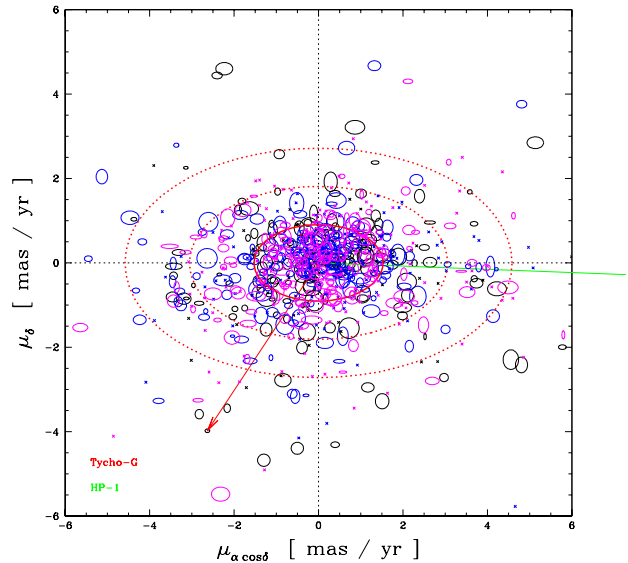}
   \caption{
Close-up view of the left panel in Fig.~\ref{VPD} around the zero motion 
of the reference frame.
The ellipse in red, with semi-axes $\sigma_{\mu_{\alpha\cos{\delta}}} = 1.38$ 
mas yr$^{-1}$ and $\sigma_\delta = 0.83$ mas yr$^{-1}$, shows the assumed PM 
distribution of objects in the field.
For reference, 
$2 \times$ and $3 \times$ larger semi-axes are shown with 
dotted lines. 
Clearly, Tycho-G has significantly different PMs compared with the bulk of 
stars in the field.
   \label{VPDzoom}
}
   \end{center}
   \end{figure}

\begin{figure}
\centering
\includegraphics[width=6.5cm,angle=+90]{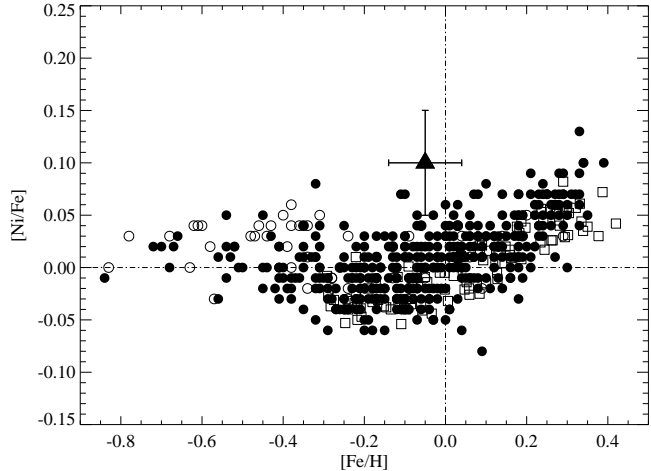}
\caption{[Ni/Fe] abundance ratio of Tycho-G (filled triangle) in 
comparison with the abundances of F-, G-, and K-type metal-rich dwarf
stars (Neves et al.\ 2009).  Thin-disk stars are depicted as filled
circles, whereas transition and thick-disk stars are the empty
circles.  Solar analogs are shown as empty squares (Gonz\'alez
Hern\'andez et al.\ 2010, 2013).  The size of the error bars indicates
the $1\,\sigma$ uncertainty.  The dashed-dotted lines indicate solar
abundance values.
\label{fabNi}}
\end{figure}

In Fig.~\ref{fabNi} we compare the [Ni/Fe] abundance ratio of Tycho-G
with those of F-, G-, and K-type stars in the solar neighbourhood from
Neves et al.\ (2009), which obviously were measured with the same
line list and thus the same log\,$gf$ values. We have corrected one of
the points at [Ni/Fe]~$=0.18$ and [Fe/H]~$=0.03$ in Neves et
al.\ (2009), which shows an unexpectedly high abundance dispersion, by
removing the Ni abundance outliers from some spectral Ni lines using a
1.5\,$\sigma$ clipping procedure (see Gonz{\'a}lez Hern{\'a}ndez et
al.\ 2013).  This gives a revised value for this star, HD~209458, of
[Ni/Fe]~$=-0.03$.  The average value of the [Ni/Fe] ratio in the
relevant range of metallicities (i.e., [Fe/H]~$= -0.05 \pm 0.09$) in
thin-disk and thick-disk stars of the sample in
Neves et al.\ (2009) is $0.00 \pm 0.03$ ($N_{\rm stars}=129$) for thin-disk
F-, G-, and K-type stars, and $0.01 \pm 0.03$ ($N_{\rm stars}=7$) for 
thick-disk F-, G-, and K-type stars. The average value for solar analogs 
in Gonz{\'a}lez Hern{\'a}ndez et al.\ (2010, 2013) is
$-0.02 \pm 0.03$ ($N_{\rm stars}=24$). The Ni abundance in Tycho-G,
taking into account all known uncertainties, still appears to be
slightly above (at almost 1.7\,$\sigma$) the Galactic trend,  
mostly defined by main-sequence stars in the solar neighbourhood (see 
Fig.~\ref{fabNi}). 
Error bars are relatively large as a result of the modest quality of the
spectrum of this very faint star, despite hours of integration time
with the 10-m Keck~I telescope.
%
%
\section{The Orbit of the Binary Precursor of SN 1572}

%

The PM measured for Tycho-G can be translated into a tangential motion
(i.e., perpendicular to the line of sight) once the distance to the
object is known. Its distance lies in the range $2.48 \pm 0.21$ kpc to
$4.95 \pm 0.52$ kpc (GH09), which encompasses the estimated distance
to the SNR ($2.83 \pm 0.79$ kpc; Ruiz-Lapuente 2004). We will assume,
in the following, that Tycho-G is indeed inside the SNR, to derive its
tangential motion.

The PM parallel to the celestial equator, $\mu_{\alpha\cos\delta} =
-2.63 \pm 0.18$ mas yr$^{-1}$, translates into $v_{\alpha} = -35 \pm
10$ km s$^{-1}$, while $\mu_{\delta} = -3.98 \pm 0.10 $ mas yr$^{-1}$
gives $v_{\delta} = -53 \pm 14$ km s$^{-1}$, where the uncertainties
come almost entirely from those in the distance. The total tangential
velocity is thus $v_t = \sqrt{v_{\alpha}^2+v_\delta^2} = 64\pm 11$ km
s$^{-1}$. Since the radial velocity of Tycho-G is $v_r = -80 \pm 0.5$
km s$^{-1}$ (GH09), its total velocity (referred to the Local Standard
of Rest) is, in absolute value, $v_{\rm tot} = \sqrt{v_t^2+v_r^2} =
102 \pm 9$ km s$^{-1}$.

We could interpret the total velocity of the star as corresponding to
that of its orbital motion and arising from disruption of the binary
orbit because of the explosion of the WD component of the system. The
average tangential velocity of disk stars at the position and distance
of the SNR is negligible compared with that measured for star
G. However, since the average radial velocity in the direction of
Tycho's SNR, at a distance $d \approx 2.8$ kpc from the Sun, is
$\langle v_r \rangle \approx -37 \pm 20$ km s$^{-1}$ (from the Besan\c
con model of the Galaxy; Robin et al.\ 2003), it would be more
appropriate to take as the peculiar radial velocity due to the orbital
motion $v_r \approx -43 \pm 20$ km s$^{-1}$. Thus, the total orbital
velocity would become only $v_{\rm orb} = 77 \pm 16$ km s$^{-1}$.
If Tycho-G were actually inside the SNR, its radial velocity would
thus be slightly more than 2\,$\sigma$ above average. Nevertheless, in
assessing the peculiarity of the kinematics of Tycho-G (which will be
done in the next section), the uncertainty in its distance must be
taken into account. Based on the same reference, the radial velocity
would be between 2.2\,$\sigma$ (lower limit on the distance; star in
the foreground of the SNR) and 1.2\,$\sigma$ (upper limit; star in the
background) above the average.
Note that the ratio of $v_t$ to $v_r$ gives the inclination of the
plane of the orbit with respect to the line of sight, and hence
determines the projection of the rotational velocity along that line
assuming coplanarity ($i \approx 34^\circ$).

Adopting 1.4 M$_{\odot}$ and 1 M$_{\odot}$ as the respective masses of
the WD and its companion at the time of the explosion, we obtain an
orbital separation of $a = (26 \pm 12)~{\rm R}_{\odot}$ and a period
of $P = 10 \pm 7$ days. Applying Eggleton's (1983) formula for the
effective Roche-lobe radius of the companion star, we have $R_{\rm L}
= (9 \pm 4)~{\rm R}_{\odot}$. This means that if Tycho-G was indeed
the companion of the SN, at the time its radius was considerably
larger than its present radius, which is within the range $R \approx
1$--2 R$_{\odot}$ (GH09).

\begin{figure}
\centering
\includegraphics[width=9cm,angle=0]{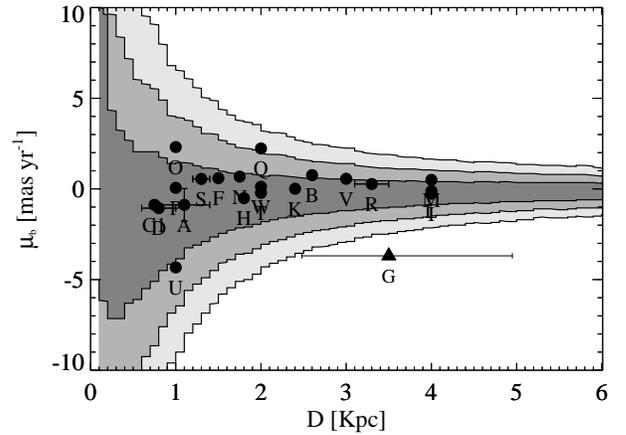}
\caption{The distribution of proper motions $\mu_{b}$, 
perpendicular to the Galactic plane, as a function of distance, in the
direction of the X-ray centroid of Tycho's SNR, for thin-disk and
thick-disk stars together (and [Fe/H] $> -0.14$), according to the
Besan\c con model of the Galaxy (Robin et al.\ 2003).  1\,$\sigma$,
2\,$\sigma$, and 3\,$\sigma$ regions are indicated.  The position of
Tycho-G is depicted as a triangle, $D [{\rm kpc}] =
3.50^{+1.45}_{-1.02}$ and $\mu_{b} [{\rm mas~yr^{-1}}] = -3.69 \pm
0.10 (0.04)$.  The range of distances to Tycho-G, as well as the PM
measured in this work, are indicated by the horizontal line.  The
lengths of the vertical segments at the two ends of the line
correspond to the error bar of the PM measurement. The positions of
all the other stars discussed in Section 9 (see also Table 3), and
with estimated distances $D \leq 6\ {\rm kpc}$, are also shown. The
error bars in $\mu_{b}$ are smaller than the sizes of the points.
\label{besancon}}
\end{figure}

\section{The Kinematics of Tycho-G and Implications}
%
%
With the present high-precision results in its proper motion, the
kinematics of Tycho-G can now be reevaluated. The new measurements
have confirmed a high proper motion perpendicular to the Galactic
plane.  As already mentioned in Section 1, a similar result is
obtained by K13.  
If the star were at the distance of Tycho's SNR 
(taken here to be 2.83 kpc), its velocity would be 
$v_{b} = -51 \pm 1.5$ km s$^{-1}$. 
Let us now separately consider the probability, for Tycho-G, of being
either a random thin-disk, thick-disk, or halo star.

\textit{Thin disk}. The metallicity and height above the Galactic
plane of Tycho-G are typical of a thin-disk star. Taking again the
Besan\c con model of the Galaxy (Robin et al.\ 2003) as an
approximation to the kinematic structure, the average for all disk
stars (thin and thick disk together) having [Fe/H] $> -0.14$, at the
Galactic latitude of the remnant and at its distance, is only $\left<
v_{b} \right> = -2.6 \pm 12$ km s$^{-1}$ (see Fig. 11). Thus, the
velocity of Tycho-G is almost 4\,$\sigma$ above the average, and the
probability of its being due to chance alone should be $P \lapprox
10^{-4}$. 
Given the limits on the distance to Tycho-G quoted in the preceding
section, we find that $v_{b}$ must, in any case, be within the range
$-45 \pm 1$ km s$^{-1}$ 
to 
$-90 \pm 3$ km s$^{-1}$. 
For the shortest possible distance
(2.48 kpc), 
$v_{b}$ would still be 
$\sim 3.7\,\sigma$ 
above average. 
If Tycho-G were in the background of the SNR, then at the upper limit
of its distance
(4.95 kpc), 
$v_{b}$ would climb to 
7.5\,$\sigma$. 
For the shortest distance, as stated in the previous section, the
radial velocity would be 2.2\,$\sigma$ above average (taking the same
model as a reference). While the metallicity of Tycho-G ([Fe/H] =
$-0.05 \pm 0.09$) is typical of a thin-disk star, [Ni/Fe] is almost
1.7\,$\sigma$ above the Galactic trend ($P \lapprox 0.1$), and thus
the total probability of finding, at random and in a single try, a
thin-disk star with such a high velocity perpendicular to the Galactic
plane and [Ni/Fe] excess together, would be the product of the two
probabilities, which gives $P \lapprox 10^{-5}$ (at the assumed
distance of the SNR), and at most $P \lapprox 10^{-4}$ (this value
corresponding to the shortest possible distance).

\textit{Thick disk}. GH09 (see their Fig. 10, and also Fuhrmann\ 2005)
have shown that there are some thick-disk stars, even within the
metallicity range of Tycho-G, having kinematics apparently similar to
that of this star (although $v_{b}$ is not shown directly in the
Toomre diagram, but only the combination $(U^{2} +
W^{2})^{1/2}$). However, at the location of Tycho's SNR, the density
of thick-disk stars with metallicities higher than the above limit is
much lower than that of thin-disk stars (in the model of reference,
the fractional density of the total of the thick-disk stars is only
3.4\%; see Table 2 of Robin et al.\ 2003). Since thick-disk stars
appear as the closest relatives of Tycho-G, we will discuss this case
more in detail. 

\begin{figure}
\centering
\includegraphics[width=85mm,angle=0]{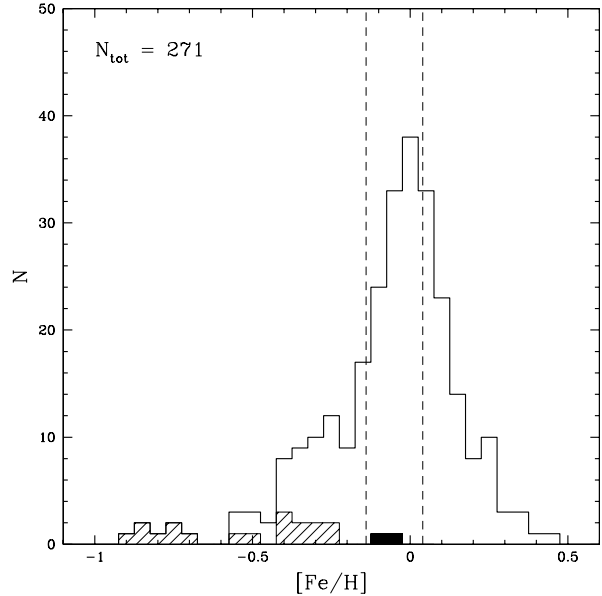}
\caption{
Metallicity distribution function of a volume--complete sample of nearby 
F, G, and K stars belonging to the Galactic disk. Shaded histogram 
corresponds to the thick--disk stars. Vertical dashed lines mark 
the metallicity range of Tycho-G. Highligted in black are the only two 
thick--disk stars with metallicities within that range. The total sample 
comprises 271 stars, and the number of stars within the said range is 128 
(adapted from Fuhrmann 2011). 
\label{fuhrmann}}
\end{figure}

The relative number density of thick-disk stars close to the Galactic
plane is somewhat uncertain: Robin et al.\ (1999) and Buser et
al.\ (1999) found values around 6\%, while Binney \& Tremaine (2008)
adopt a value as low as 2\%.  Concerning metallicity, although the
average metallicity of thick-disk stars is lower than that of
thin-disk stars, there is, nonetheless, some overlap. However, the
fraction of thick-disk stars with metallicities [Fe/H] $> -0.14$ is
very small.  In the same Fig. 10 of GH09, for instance, a sample of 38
thick-disk stars reduces to only 6 when the metallicity constraint is
applied. This would mean that such stars (taking 6\% as the fractional
density of thick-disk stars) constitute only $\sim 0.9$\% of the
number density in the region where Tycho-G is found. In a
volume-complete sample of thin-disk and thick-disk stars in the solar
neighbourhood (a total of 271 F-, G-, and K-type stars), 
Fuhrmann (2011, Fig. 15) finds only 2 thick-disk stars in the
metallicity range of Tycho-G, giving a relative number density of
$\sim 0.7$\% for these ``transition stars.''
However, as we show in Figure~\ref{fuhrmann}, the number of stars in the
sample that have metallicities within the range of 
that of 
Tycho-G is only
128, so the relative number density of ``transition stars'' in that
range becomes $\sim$ 1.5\%.

If we were to adopt $\sigma_{W}$ = 39 $\pm$ 4 km s$^{-1}$ (Bensby et
al. 2003) for the velocity dispersion of thick--disk stars
perpendicular to the Galactic plane, we would find $v_{b}$ of Tycho-G
(at the distance of the SNR) to be 1.28\,$\sigma$ above the average
thick--disk stars ($P \lapprox 0.2$).
Multiplying by the probability of finding, at random, a thick--disk
star with a metallicity as high as that of Tycho-G, 0.015, as we have
just seen, we would have $P\lapprox 0.003$.  However, the dispersion
in $v_{b}$ of the thick--disk stars with metallicities as high as that
of Tycho-G is smaller.
Using again the Besan\c con model of the Galaxy, we find that for
those stars with a velocity dispersion of only $\sigma_{W} \simeq$ 20
$\pm$ 2 km s$^{-1}$ (intermediate between that of the bulk of thick--
and thin--disk stars), the $v_{b}$ of Tycho-G would be slightly above
2.5\,$\sigma$ ($P \lapprox 0.012$) if it were at the distance of the
SNR.
For the nearest possible distance, since then $v_{b}$ is only -45 km
$^{-1}$, it would be at 2.25\,$\sigma$ ($P \lapprox 0.024$), and for
the largest distance ($v_{b}$ = -90 km s$^{-1}$), it would be at
4.5\,$\sigma$. Therefore, the probability of finding, at random, a
thick--disk star with a metallicity above the lower limit for that of
Tycho-G, and moving at least as fast, within the range of possible its
distances, becomes $P \lapprox 3.7\times10^{-4}$ at most (star at a
distance of only 2.48 kpc).

In the preceding discussion, we have not yet taken into account any Ni
overabundance. We see from Section 4 that the Ni abundance ratio
[Ni/Fe] = $0.10 \pm 0.05$ is 1.3\,$\sigma$ above the Galactic trend
for thick-disk stars ($P \lapprox 0.2$), and the same reasoning as for
thin-disk stars applies: the probability of finding by chance a
thick-disk star with a metallicity as high as that of Tycho-G, and
with its kinematics and Ni overabundance, in a single try, is $P \leq
7.4\times10^{-5}$ (i.e., less than 1 in $\sim$\, 13500).

\textit{Halo}. Although the value of $v_{b}$ for Tycho-G falls within
the velocity dispersion for halo stars (85 km s$^{-1}$, according to
Robin et al.\ 2003), the possibility that it belongs to such a
population can be almost completely excluded. Not only is the number
density of halo stars at the position and range of distances of
Tycho-G $\sim 10^{-3}$ times the local density, but for those stars
[Fe/H] = $-1.5 \pm 0.5$ (Gilmore \& Wyse 1985), which means that the
metallicity of Tycho-G ([Fe/H] = $-0.05 \pm 0.09$) is almost
3\,$\sigma$ above the average. The probability of finding such a star
at random is thus $P \lapprox 3 \times 10^{-6}$ (combining metallicity
and relative star density).

If Tycho-G were just a chance interloper, it would most likely be a
thick-disk star. As we have seen, this means having picked a star,
unrelated to the SNR but within its distance range and in a small area
around its centre, of which there are fewer than 
1 in $\sim 13500$.  
However, as discussed in Section 9, there are four other stars,
apart from Tycho-G, at distances at least marginally compatible with
that of the SNR and inside the explored area of the sky, which would
increase the probability by a factor of 5 
($P \lapprox 3.7\times10^{-4}$).

The low rotational velocity of Tycho-G ($v_{\rm rot}\,{\rm sin}\,i
\lapprox 6.6$ km s$^{-1}$; see GH09) is well explained even if we
assume that the rotational period $P_{\rm rot}$ has remained unchanged
after the explosion and that it was $P_{\rm rot} = P_{\rm orb}$
(synchronous rotation). From GH09, the current radius of the star must
be $R \approx 1$--2 R$_{\odot}$.  With a period $P_{\rm rot} \approx
10$ days (from the preceding section), and for the larger radius,
$v_{\rm rot} \approx 10$ km s$^{-1}$; with $i \approx 34^\circ$ (the
value obtained from our reconstruction of the orbit, Section 5),
$v_{\rm rot}\,{\rm sin}\,i \approx 5.6$ km s$^{-1}$, a small
value. The effects of the impact of the SN ejecta on the companion
studied by Pan et al.\ (2012a) and Liu et al.\ (2013) would reduce
$v_{\rm rot}$ even more. Concerning this last point, it should be
noted that the hydrodynamic simulations have been done only for
main-sequence companions, and that the removal of angular momentum
from less compact stars should be larger.

In conclusion, there is a very low probability ($P \lapprox 0.00037$)
for a star with the kinematics, metallicity, and Ni excess of Tycho-G,
and within the distance range of Tycho's SNR, to have been found by
chance when exploring the central region of the remnant.

~\\


\section{The Position of Tycho-G}

The position of Tycho-G is within the uncertainties pertaining to the
location of the site of the explosion inside SNRs.  In the case of
Tycho, the centroids of the X-ray emission determined from the {\it
  Chandra X-ray Observatory} images (Warren et al.\ 2005) and from
{\it ROSAT} (Hughes 2000) differ by only $6.5''$, but they are $\sim
26$--$28''$ away from the centroid of the radio emission, determined
from VLA observations (Reynoso et al.\ 1997). The current position of
Tycho-G (see Table 2) is at an angular distance of $29.8''$ from the
{\it Chandra} X-ray centroid ($\alpha_{\rm J2000.0} = 00^{\rm
  h}25^{\rm m}19.40^{\rm s},\ \delta_{\rm J2000.0} =
+64^{\circ}08'13.98''$), which amounts to $\sim 10$\% of the radius of
Tycho's SNR. From the PM obtained in Section 3, the position of the
star in 1572 was $\alpha_{\rm J2000.0} = 00^{\rm h}25^{\rm
  m}23.75^{\rm s},\ \delta_{\rm J2000.0} = +64^{\circ}08'03.73''$, and
its angular distance with respect to the same centroid should thus
have been similar, $30.3''$.

Asymmetries can arise from interaction of the SNR with the ambient
medium, but they can also be intrinsic to the SN ejecta (and, of
course, both can occur simultaneously).
In Tycho, there is evidence that the ejecta encountered a dense H
cloud at the eastern edge, giving rise to brighter emission and lower
velocity there (Decourchelle et al.\ 2001), which would place the site
of the explosion to the E of the centroid of the SNR.
More recently, Williams et al.\ (2013), from mid-infrared observations
of dust emission, find an overall gradient in the ambient density,
with densities being higher in the NE than in the SW. From
two-dimensional hydrodynamic simulations, they find that an overall
round shape of the SNR is produced, but that the centre of the
explosion is then offset from the geometric centre by $\sim 10$\% in
the direction of the higher ambient density (that is, toward the
NE). At the same time, they favor a distance of 3--4 kpc to the SNR.
The second kind of asymmetry (intrinsic) has been found by Winkler et
al.\ (2005) in the remnant of SN 1006, another SN~Ia. The projection
of the remnant on the plane of the sky looks quite round, as in the
case of Tycho, but observations of background UV sources show that the
Fe-rich ejecta are egg-shaped, and elongated at an angle with the line
of sight; see Figure 8 of Winkler et al.\ (2005), and also their
Figure 1 for the projection on the plane of the sky. This work has
recently been extended to the distribution of the O-burning and
incomplete Si-burning products (Si, S, and Ar) by Uchida et
al.\ (2013), who deduce a velocity asymmetry of $\sim 3100$ km
s$^{-1}$. Winkler et al.\ (2005) concluded that the position on the
sky of the site of the explosion differs from the X-ray centroid by as
much as $\sim 20$\% of the radius of the SNR. Kerzendorf et
al.\ (2012) have objected that the models of asymmetric SN~Ia
explosions by Maeda et al.\ (2010) confine the asymmetry to the
distribution of the innermost ejecta, while the outer ejecta show
spherical symmetry, and that the position of the centre of the WD, at
the time of the explosion, should rather be given by the centroid of
the outer ejecta.  However, in those models, a velocity offset of the
inner ejecta (with respect to the original centre of the star) is
attributed to the deflagration phase of the explosion, while the more
spherically symmetric mass ejection is produced by the transition from
deflagration to detonation, closer to the surface. It is hard to
imagine that while the inner layers are moving away from the original
centre, the centroid of the outer layers remains anchored in that
position. Thus, in this type of asymmetric explosion, neither the
present centroid of the Fe-rich ejecta nor that of the outer ejecta
should correspond to the position of the centre of the WD at the time
of the explosion.
No similar analysis has been done for SN 1572, but the spectrum of its
light echo suggests that the explosion was also aspherical (Krause et
al.\ 2008).

Among the stars with $m_{V} \leq 22$ mag, the closest one to the
Chandra X-ray centroid is Tycho-B (see Fig.~\ref{tychofield}, and
Tables 2 and 3).  In order of increasing distance are stars Tycho-A,
-C, -E, -D, -F, -J, and then -G.  As discussed in Section~9, none of
these other stars shows any sign of being related to the SN explosion.

\section{The Luminosity of Tycho-G} 

The luminosity of Tycho-G is within the range $1.9 < L_{*}/{\rm
  L}_{\odot} < 7.6$, as calculated from its effective temperature and
surface gravity (assuming a mass of 1 M$_{\odot}$; GH09). What should
be expected for the companion of a SN as recent as SN 1572?

Podsiadlowski (2003) calculated the luminosity evolution of a subgiant
star of 1 M$_{\odot}$, $R_{*} = 2.5~{\rm R}_{\odot}$, and $L_{*}
\approx 3~{\rm L}_{\odot}$, after being hit by the ejecta of a
SN~Ia. He assumed that 0.2 M$_{\odot}$ were removed by the impact
(based on the hydrodynamic calculations of Marietta et al.\ 2000), and
that variable amounts of energy were deposited uniformly in the
outermost 90\% of the radial extent of the remaining object
(containing 0.57 M$_{\odot}$). Such amounts ranged from $4 \times
10^{47}$ erg (close to the energy needed to unbind those layers
completely) down to zero. He found, in all cases, that the luminosity
evolution was initially much faster than the Kelvin-Helmholtz time
scale of the presupernova subgiant, since it was determined by the
thermal time scale of the outer layers of the star, which is many
orders of magnitude shorter.  Depending on the amount of energy
deposited, the luminosity of the companion, 440 yr after the SN
explosion, might range from $\sim 200~{\rm L}_{\odot}$ to $\sim
0.1~{\rm L}_{\odot}$ (see Fig. 2 in Podsiadlowski 2003). The result
also depends strongly on the mass of the star at the time of the
explosion and on the amount of mass removed by the explosion (which,
in turn, depends on the evolutionary stage of the subgiant).

Recently, Shappee et al.\ (2013) have calculated the post-explosion
evolution of a main-sequence companion of 1 M$_{\odot}$. From the
hydrodynamic simulations of Pan et al.\ (2012a), which show that in
this case most ($\sim 65$\%) of the mass lost from the impact of the
ejecta is due to ablation (heating by the shock front) and the rest
($\sim 35$\%) is removed by stripping (momentum transfer by the
shock), they construct an initial model by heating the whole star
until it generates a stellar wind able to expel the same amount of
mass that is lost by ablation in the hydrodynamic calculation. They
then follow the evolution of the star by means of a standard
one-dimensional code, finding that after $\sim 500$ yr, the luminosity
of the companion should still be $\geq 20~{\rm L}_{\odot}$, while
$T_{\rm eff} \leq 5500$ K. Note that their pre-explosion model has a
radius $R \approx 1~{\rm R}_{\odot}$, significantly smaller than that
inferred in Section 5 for Tycho-G at the same stage under the
assumption that it was the companion of SN 1572.

Pan et al.\ (2012b), based on their own hydrodynamic models (Pan et
al.\ 2012a), find that the evolution of the remnant star strongly
depends not only on the amount of energy absorbed from the explosion,
but also on the depth of the energy deposition. They calculate the
evolution of several pre-explosion models through the hydrodynamic and
hydrostatic stages. One of them (their model E), 440 yr after the
explosion, is the closest in mass, radius, and effective temperature
to Tycho-G, having a luminosity higher than that of Tycho-G by only a
factor of two. Again, however, the initial model is more compact than
Tycho-G should have been at the time of the explosion.
As it can be seen from the hydrodynamic simulations of Marietta et
al.\ (2000), the less compact the star is, the more mass is removed by
stripping and less by ablation, which also means less energy deposited
into the layers that remain bound.

We stress once again that the calculations that predict, for a stellar
SN~Ia companion 440 yr after the explosion, luminosities higher than
that of Tycho-G, have all been made for main-sequence stars, and that
the outcomes might be significantly different for even slightly more
evolved ones (a point that also affects rotation, as indicated in
Section 6). The reconstruction of an evolutionary scenario, leading to
a pre-explosion configuration similar to the one deduced in Section 5,
should include nonconservative mass transfer and the different
evolution of the orbital separation before and after inversion of the
mass ratio between the two stars. We also note that, depending on the
amount of mass lost from the impact of the SN ejecta, on the amount of
energy injected, and on how that energy is distributed inside the
remaining star, the object might not have reached thermal equilibrium
yet; it could be slowly expanding on a Kelvin-Helmoltz time scale,
obtaining energy for this from internal sources, as suggested by
Podsiadlowski (2003).


\section{Other Stars around the Centre of Tycho's SNR}

\begin{figure}
\centering
\includegraphics[width=85mm,angle=0]{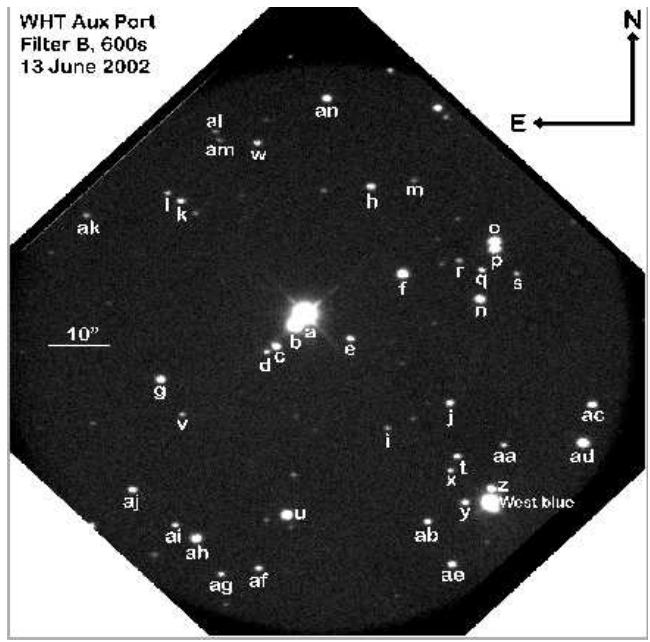}
\caption{
$B$-band image taken with the 4.2-m William Herschel telescope,
  showing all of the named stars near the centre of Tycho's SNR.
\label{tychofield}}
\end{figure}

In RL04, radial velocities were measured and distances estimated for
13 stars (including Tycho-G) located within an angular distance of
$39''$ from the centroid of the X-ray emission of Tycho's SNR and with
$V < 22$ mag, as well as for 3 slightly more distant stars, up to
$41.5''$ from the same point and also brighter than the said magnitude
(Tables 1 and S3 of RL04). Although PMs measured with WFPC2 aboard
{\it HST} were graphically displayed in Figure~1 of RL04, they were
numerically given only for Tycho-G. Now, with the much more precise
PMs in Table 2 of the present paper, and with new estimates of the
distances for some of those stars, we can rediscuss the kinematics of
the sample and the possible association of each of its members with SN
1572. In Table 3 we give $BVR$ photometry, estimated distances, and
velocities $v_{\alpha}$ (parallel to the celestial equator) and
$v_{\delta}$ (perpendicular to it) for the full set of 24 stars
located within $42''$ from the X-ray centroid (see
Fig.\ref{tychofield} for labels and positions) and brighter than $V =
22$ mag. We briefly comment on each of them next.

Tycho-A is a giant star, at a distance between $1.1 \pm 0.3$ kpc
(RL04) and $1.4 \pm 0.8$ kpc (K13), thus closer than Tycho's SNR. Its
radial velocity $v_r$ (referred to the Local Standard of Rest, LSR) is
between $-23$ and $-28.5$ km s$^{-1}$.  Owing to saturation even in
the {\it HST}/WFC3/UVIS short images, no PM has been measured in the
present work; however, K13 (who use 0.5~s ACS/WFC exposures to derive
their master frame) give $\mu_{\alpha cos \delta} = -0.09 \pm 1.17$
mas yr$^{-1}$ and $\mu_{\delta} = -0.89 \pm 0.90$ mas yr$^{-1}$, which
translate into $v_{\alpha} = -0.5 \pm 3$ km s$^{-1}$ and $v_{\delta} =
-4.6 \pm 5$ km s$^{-1}$ assuming the RL04 distance. The radial
velocity is thus consistent with the distance for a star belonging to
the Galactic thin disk, and the small total tangential velocity is
also typical of such a stellar population.

Tycho-B is a main-sequence star of spectral type A8--A9, at a distance
between $2.6 \pm 0.5$ kpc (RL04) and $1.8 \pm 0.8$ kpc (K13).  Its
radial velocity is between $-38$ and $-44.5$ km s$^{-1}$, indicating
that Tycho-B is a thin-disk object, like Tycho-A; its higher radial
velocity corresponds to the larger distance. The components of the
tangential velocity, calculated from the PMs in Table 2 ($v_{\alpha}
\approx -21$ km s$^{-1}$, $v_{\delta} \approx 7$ km s$^{-1}$), are
again typical of its population. It should be noted that the high
rotational velocity measured by K13, $v_{\rm rot}\,{\rm sin}\,i =
171^{+16}_{-33}$ km s$^{-1}$, is well within the range covered by
stars of its spectral type (Abt \& Morrell 1993, 1995).
Thompson \& Gould (2012) have attempted to relate Tycho-B to the SN, based on the 
low metallicity found by K13, by making it a member of a quadruple system: 
one binary pair giving a double WD system that produces the SN, 
plus another pair merging into a blue straggler (Tycho-B). The problem here
is that the kinematics of Tycho-B very precisely match that of a thin-disk
star. 

Tycho-C corresponds to two different objects: a brighter, bluer star
(C1) with $V = 19.06$ mag and $B-V = 2.00$ mag, and a fainter, redder
one (C2) with $V = 20.53$ mag and $B-V = 2.38$ mag.
 
C1 appears to be a main-sequence star of spectral type K7 in the
foreground of the SNR. It is at a distance of less than 1 kpc. Its
radial velocity ($-33 \pm 6$ km s$^{-1}$) was measured by RL04.  Its
PM (Table 2) gives the two components of its tangential velocity
(Table 3) as $v_{\alpha} \approx -8$ km s$^{-1}$, $v_{\delta} \approx
-0.4$ km s$^{-1}$, both of which are quite small.  The star was
classified by K13 as a red giant at $d = 5.5 \pm 3.5$ kpc, instead;
however, from its colour, with a reddening $E(B-V) = 0.76$ mag (GH09)
or $E(B-V) = 0.86$ mag (estimated by K13 for Tycho-B, at a much
smaller distance), star C1 would be of spectral type K2, with $M_{V}
\approx +0.2$ mag, and then its apparent magnitude would place it at
an even larger distance, which is inconsistent with the small radial
velocity.
 
C2 is probably a red-giant star in the background.  No reliable
measurement of $v_r$ has been possible, but the PM was determined
(Table 2). The combination of distance and PM makes it a member of the
halo population.

Tycho-D is on the main sequence and has spectral type M1 (RL04; GH09).
It is at a distance less than 1 kpc. No radial velocity could reliably
be measured by RL04, but K13 give $v_r = -50.6 \pm 0.8$ km s$^{-1}$,
which would be high for such a small distance (perhaps a hint of
binarity). We see that both components of the tangential velocity are
small ($v_{\alpha} \approx -8$ km s$^{-1}$, $v_{\delta} \approx -5$ km
s$^{-1}$).

Tycho-E is a double-lined spectroscopic binary (GH09). Treated as a
single star, it appears as a K2--K3 giant at a large distance, with
$v_r = -26 \pm 18$ km s$^{-1}$ (RL04) or $v_r = -55.91 \pm 0.27$ km
s$^{-1}$ (K13). The discrepancy is likely to come from the binarity,
the star having been observed at different orbital phases in each of
the two studies.  The same might account for the claim by Ihara et
al.\ (2007) that this star shows blueshifted iron absorption
lines. For $d > 20$ kpc, $v_{\alpha} \gapprox -95$ km s$^{-1}$ and
$v_{\delta} \gapprox +27$ km s$^{-1}$, not particularly high for a
star located at least $\sim 0.5$ kpc above the Galactic plane.

Tycho-F is on the main sequence, at spectral type F9 (GH09). From
photometry, its distance should thus be $\sim 1.5$ kpc. The radial
velocity is $v_{r} = -34 \pm 11$ km s$^{-1}$ (RL04). From its proper
motion, both components of its tangential velocity are small
($v_{\alpha} \approx -23.5$ km s$^{-1}$, $v_{\delta} \approx 2$ km
s$^{-1}$).

Tycho-G has been extensively discussed already. In Table 3 we show the
two most extreme estimates of its distance (GH09) and the
corresponding components of the tangential velocity.

Tycho-H, from its colour, could be either a giant of spectral type
K0--K1 at a distance $\sim 24$ kpc, or a main-sequence star of the
same spectral type at $d \approx 1.8$ kpc. Its radial velocity is
$v_{r} = -71 \pm 10$ km s$^{-1}$, in better agreement with the longer
distance. From the PMs, either $v_{\alpha} \approx -360$ km s$^{-1}$
and $v_{\delta} \approx -96$ km s$^{-1}$ (large distance), or
$v_{\alpha} \approx -27$ km s$^{-1}$ and $v_{\delta} \approx 7$ km
s$^{-1}$ (short distance). The star would belong to the halo in the
first case and to the thin disk in the second one.  For the short
distance the radial velocity would be unusually high, but not the
tangential velocity.

Tycho-I, from its colours (Table 3), could be either a K1--K2
main-sequence star at a distance $\sim 4$ kpc or a K6--K7 red giant at
a very large distance. Its PMs have been determined (see Table 2), and
we can thus calculate the components of its tangential velocity. In
the first case, the tangential velocity would be small ($v_{\alpha}
\approx 13$ km s$^{-1}$, $v_{\delta} \approx 4$ km s$^{-1}$).

Tycho-J is on the main sequence, with spectral type G8, at a distance
$\sim 9$ kpc. Its radial velocity comes from RL04, and its PM gives,
for the two components of its tangential velocity, $v_{\alpha} \approx
-100$ km s$^{-1}$, $v_{\delta} \approx -12$ km s$^{-1}$.

Tycho-K could be either a G9--K0 giant or a G9 main-sequence star.  In
the first case, its distance would be $\sim 27$ kpc, and in the second
case only $\sim 2.4$ kpc. Its radial velocity would fit well with the
shorter distance. On the other hand, the PM is small, in better
agreement with the longer distance.

Tycho-L, from the colours, should be of spectral type K0--K1 and could
be, like the other stars in the sample for which only photometry is
available, either on the main sequence or on the red-giant branch.  In
the former case, its distance would be $\sim 4$ kpc. In the second
case, star L would be tens of kpc away. The components of the
tangential velocity given in Table 3 ($v_{\alpha} \approx 7$ km
s$^{-1}$, $v_{\delta} \approx 1.5$ km s$^{-1}$) are those
corresponding to the shorter distance.

Tycho-M, from Table 3, could be either a main-sequence star of
spectral type K2 or a red giant of the same type. In the first case,
its distance would be $\sim 4$ kpc; thus, from the PMs in Table 2, the
components of the tangential velocity would have the small values
given in Table 3 ($v_{\alpha} \approx 12$ km s$^{-1}$, $v_{\delta}
\approx 8$ km s$^{-1}$). If it were a red giant, it would be very far
in the background.

Tycho-N is a main-sequence star of spectral type G0--G2, at a distance
between 1.5 and 2 kpc. Its radial velocity comes from RL04. The
kinematics are standard.

Tycho-O had its radial velocity measured by RL04.  From its colour, it
is a G5 main-sequence star, at $d < 1$ kpc.  If it were instead a G5
giant, its distance would be $\sim 7$ kpc.  In Table 3, the upper
limits to the components of the tangential velocity are given for the
shorter distance.

Tycho-P has a colour indicating a G2--G3 main-sequence star at a
distance $\sim 1$ kpc. Were it a G4 giant, its distance would be $\sim
8$ kpc.

Tycho-Q is either a K3 main-sequence star at a distance $\sim 2$ kpc
or a K2--K3 giant at a very large distance.  $v_{\alpha}$ and
$v_{\delta}$ in Table 3 are given for the former case.

Tycho-R is either a K2 main-sequence star at $d = 3.3 \pm 0.2$ kpc or,
again, a very distant red giant of the same spectral type. As in
previous cases, the tangential velocities are given only for the
shorter distance.

Tycho-S is either a K8 main-sequence star at $d = 1.3 \pm 0.1$ kpc or
a very distant M1 giant.

Tycho-T has an unknown radial velocity.  From its colour,
corresponding to K1, the distance could be either $\sim 2$ kpc if a
main-sequence star or $\sim 30$ kpc if a red giant. In Table 3 we give
the components of its tangential velocity only for the shorter
distance.

Tycho-U has $v_{r} = -38 \pm 4$ km s$^{-1}$ (RL04). It is a G0
main-sequence star at a distance $\sim 1$ kpc. If it were a giant
star, its distance would be $> 8$ kpc.

Tycho-V has no measured radial velocity. It appears to be a
main-sequence star of spectral type K3--K4, at a distance $\sim 3$ kpc.

Tycho-W is a K3 main-sequence star at $\sim 2$ kpc.

From Table 3, we see that only four stars (B, K, R, and V), apart from
Tycho-G, are at distances even marginally compatible with that of
Tycho's SNR. None of them shows the slightest kinematic peculiarity
that might suggest any link with SN 1572.  We also note that within
this small sample of 23 stars, none shows velocities perpendicular to
the Galactic plane comparable with that of Tycho-G, in full agreement
with our previous discussion. Given the small number of stars in the
central region of Tycho's SNR, it is unlikely that one of them would
show the unusual characteristics of Tycho-G without being related to
the SN explosion.

%
\section{Summary and Conclusions}
%

A very accurate determination of the proper motions of 872 stars in
the central region of Tycho's SNR, based on images taken with {\it
  HST} in up to four different epochs and spanning a total of 8 years,
has confirmed a high PM of Tycho-G perpendicular to the Galactic
plane: $v_{b} = -50 \pm 14$ km s$^{-1}$. The probability of its being
caused by chance alone is very small.

From the PM plus the radial velocity of Tycho-G, we have deduced the
orbital separation, the orbital period, and the orientation of the
plane of the orbit at the time of the explosion, if its present
velocity comes from the orbital motion when Tycho's SN exploded. We
also derive the radius it should have had if it were then filling its
Roche lobe. This radius is significantly larger than the current
radius, and we speculate that the star might now be expanding on a
thermal time scale.

We have also recalculated the Ni abundance of Tycho-G using the same
high-resolution spectrum as in the previous work by GH09, but this
time with an automated procedure to search for and fit the continuum
and to measure equivalent widths. We find a ratio [Fe/Ni] $= 0.10 \pm
0.05$. Compared with the Galactic trend given by new, high-quality
data on F-, G-, and K-type metal-rich dwarf stars, there is an
overabundance, although only at the 1.7\,$\sigma$ level.

The significance of the kinematics of Tycho-G is evaluated from the
new data. We discuss the probability of its being a metal-rich
thick-disk star with a high velocity perpendicular to the Galactic
plane and also with a high Ni abundance, and find it to be 
$P \lapprox 0.00037$. 
Given the parameters of the orbit and the present radius of the star,
its low rotational velocity is well explained, even without
considering the removal of angular momentum by the impact of the
ejecta.

The position of Tycho-G with respect to the centroid of the X-ray
emission of Tycho's SNR is within the uncertainties affecting the
location of the site of the explosion in other SNRs (see also the
discussion in RL04).  The star's luminosity is close to the
predictions of Pan et al.\ (2012b), for main-sequence companions of
SNe~Ia at the age of Tycho.  Tycho-G, however, should have already
started to evolve away from the main sequence at the time of the
explosion.

There is not, at present, any solid argument against Tycho-G being the
surviving companion of SN 1572. On the contrary, its kinematic
characteristics and distance range strongly point to this conclusion,
even in the absence of any Ni enhancement. The alternative
possibility, that it is just an interloper found by pure chance has a
very low probability.  From our survey of the central region of the
SNR, we conclude that, apart from Tycho-G, there is no possible
candidate for the SN companion.  Therefore, if Tycho-G were to be
discarded by some future findings, that would favor a
double--degenerate system as the origin of SN 1572, as has been
plausibly found for SN 1006 (Gonz\'alez Hern\'andez\ et al.\ 2012;
Kerzendorf et al.\ 2012) and for the progenitor of SNR 0509--67.5 in
the LMC (Schaefer \& Pagnotta 2012).

Most recent searches for SN~Ia companions have either shown their
absence or put strong limits on their presence, thus tilting the
balance toward the double-degenerate channel. From the metallicity
distribution of G-type dwarfs, however, contributions from both the
single-degenerate and the double-degenerate channels appear to be
necessary (Mennekens et al.\ 2012). Based on our study, SN 1572
appears to be a very good candidate for the single-degenerate channel.

Hydrodynamic simulations of the impact of SN~Ia ejecta on companion
stars having already left the main sequence, as well as calculations
of their subsequent evolution, are required for additional progress in
this field.



\section{Acknowledgments}

This work is based on observations with the NASA/ESA {\it Hubble Space
  Telescope}, obtained at the Space Telescope Science Institute
(STScI), which is operated by AURA, Inc., under NASA contract NAS
5-26555; the specific programs were GO-9729, GO-10098, and GO-12469.
P.R.-L.\ and R.C.\ acknowledge support from grant AYA2009--13667,
financed by the MICINN of Spain.  J.I.G.H.\ received financial support
from the Spanish Ministry project MINECO AYA2011-29060, and also from
the Spanish Ministry of Economy and Competitiveness (MINECO) under the
2011 Severo Ochoa Program MINECO SEV-2011-0187.  A.V.F.\ is grateful
for the support of the Christopher R. Redlich Fund, the TABASGO
Foundation, and NSF grants AST-0908886 and AST-1211916; funding was
also provided by NASA grants GO-10098, GO-12469, and AR-12623 from
STScI. Some of the data presented herein were obtained at the
W.\ M.\ Keck Observatory, which is operated as a scientific
partnership among the California Institute of Technology, the
University of California, and NASA; the observatory was made possible
by the generous financial support of the W.\ M.\ Keck Foundation.  We
thank the referee, Wolfgang Kerzendorf, for his careful reading of the
manuscript and for useful suggestions that improved our work.

\noindent

%

%

\begin{table*}
\begin{center}
\includegraphics[angle=90, height=220mm]{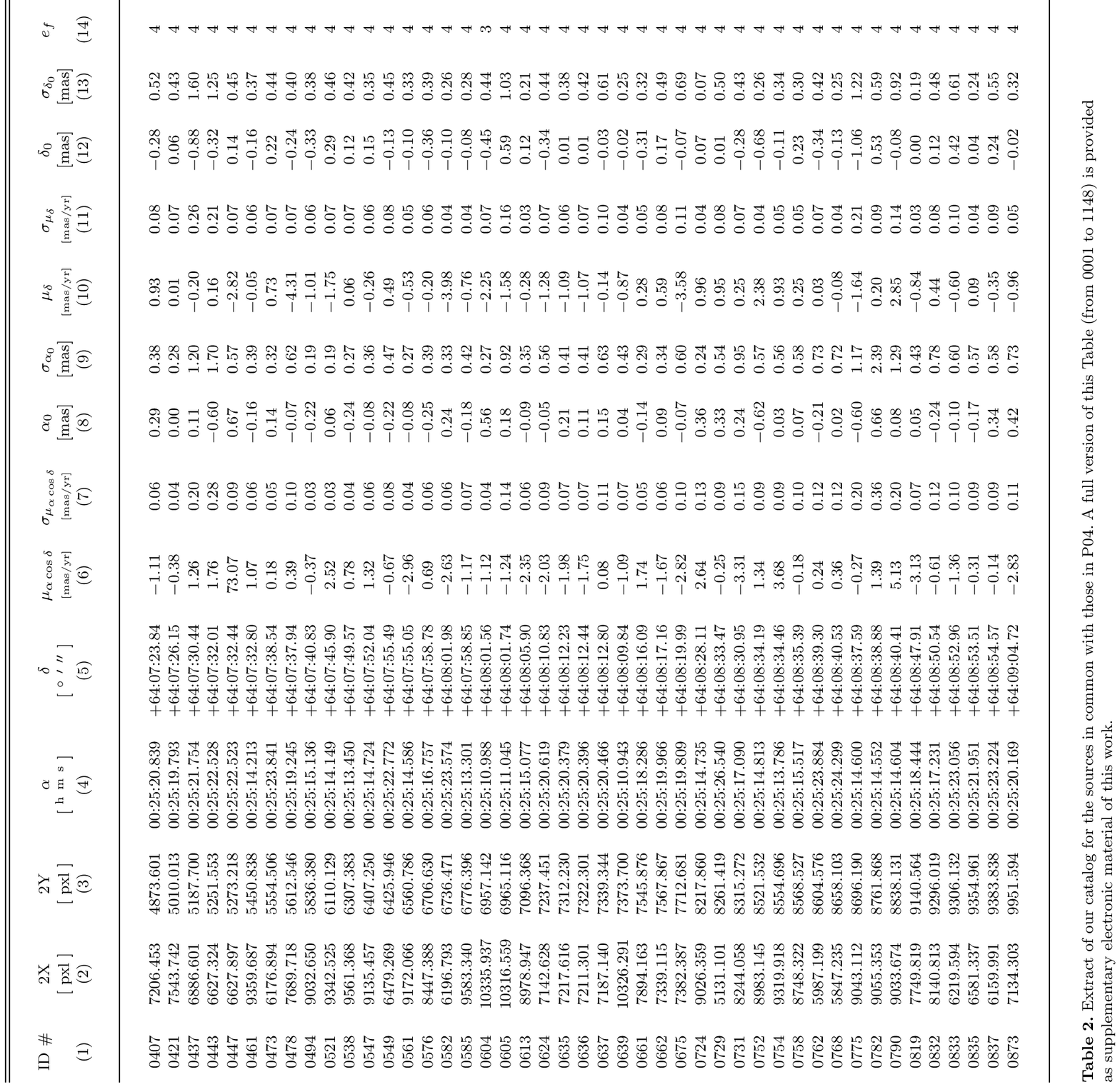}
\end{center}
\end{table*}

\begin{table*}
\begin{center}
\includegraphics[angle=90, height=220mm]{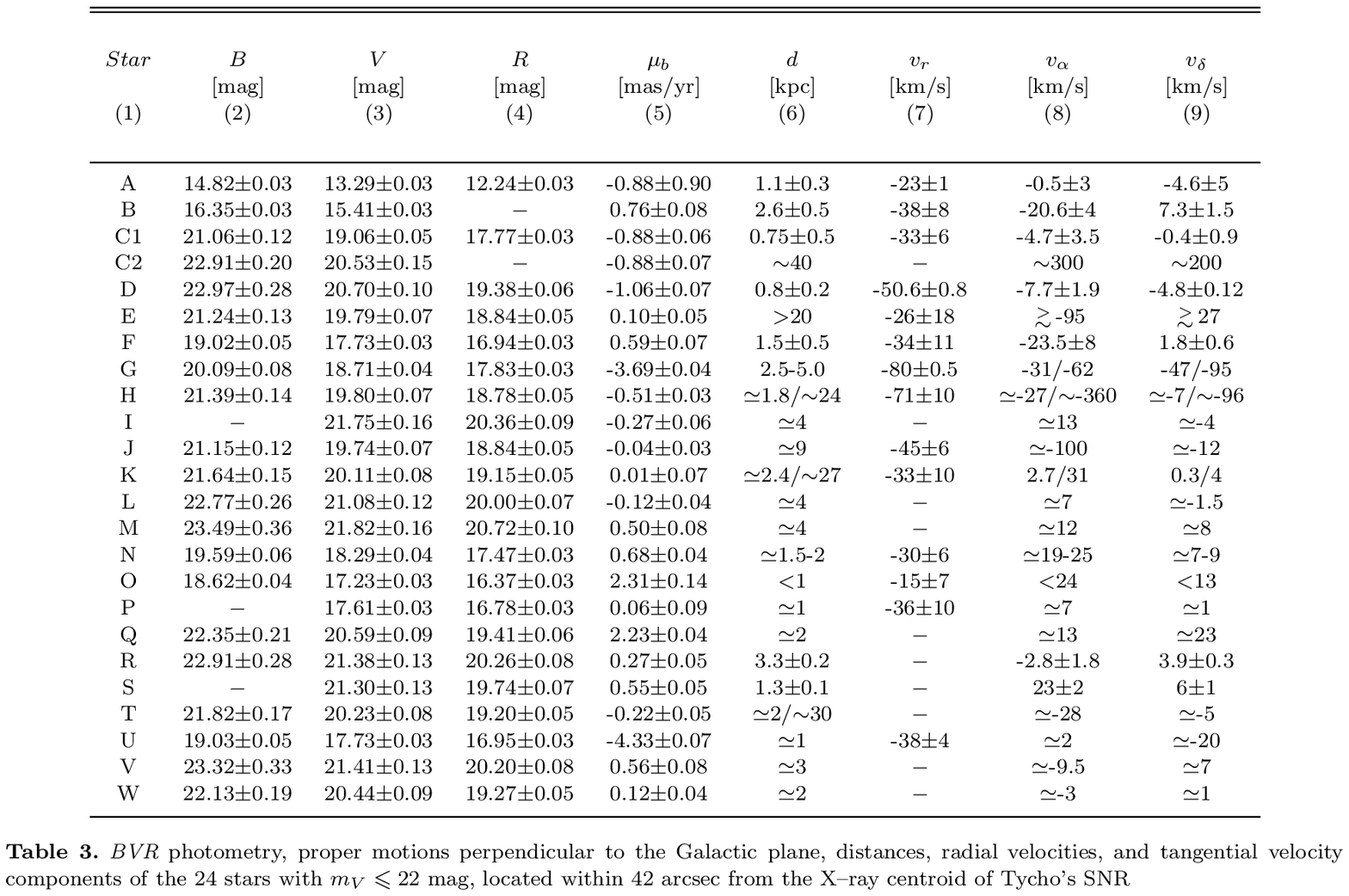}
\end{center}
\end{table*}

\label{lastpage}


\end{document}